\documentclass[runningheads]{llncs}

\usepackage{geometry}
\usepackage{graphicx}
\usepackage{amsmath,amssymb} 
\usepackage{lineno}
\usepackage{color}
\usepackage{bussproofs}

\usepackage{times}
\usepackage{epsfig}
\usepackage{eufrak}

\DeclareMathOperator{\st}{s.t.}

\DeclareMathOperator{\BOOL}{bool}

\DeclareMathOperator{\UNIT}{unit}

\DeclareMathOperator{\TRUE}{true}
\DeclareMathOperator{\FALSE}{false}
\DeclareMathOperator{\IF}{if}
\DeclareMathOperator{\THEN}{then}
\DeclareMathOperator{\ELSE}{else}
\DeclareMathOperator{\LET}{let}
\DeclareMathOperator{\IN}{in}
\DeclareMathOperator{\NORM}{Norm}
\DeclareMathOperator{\TYPES}{Types}
\DeclareMathOperator{\REDUCTIONS}{Reductions}
\DeclareMathOperator{\RANDOM}{random}
\DeclareMathOperator{\OBSERVE}{observe}
\DeclareMathOperator{\NOTYPE}{no-type}
\DeclareMathOperator{\NF}{NF}

\begin{document}


\mainmatter
\pagestyle{plain}

\title{A Probabilistic Dependent Type System based on Non-Deterministic \\ Beta Reduction (Preprint)} 


\author{Jonathan H. Warrell\inst{1}\inst{2}$^*$}

\institute{{\em Gene Expression and Biophysics group, Council for Scientific and Industrial Research, Pretoria, South Africa}
\and {\em Division of Chemical Systems and Synthetic Biology, Faculty of Health Sciences, University of Cape Town, South Africa} \\ \quad \\ *jonathan.warrell@gmail.com}
\maketitle

\begin{abstract}
We introduce {\em Probabilistic Dependent Type Systems} (PDTS) via a functional language based on a subsystem of intuitionistic type theory including dependent sums and products, which is expanded to include stochastic functions.  We provide a sampling-based semantics for the language based on non-deterministic beta reduction.  Further, we derive a probabilistic logic from the PDTS introduced as a direct result of the Curry-Howard isomorphism.  The probabilistic logic derived is shown to provide a universal representation for finite discrete distributions.
\end{abstract}

\section{Introduction}\label{sec:intro}

Non-deterministic beta reduction has been used to provide probabilistic semantics for stochastic functional languages based on the untyped lambda calculus \cite{goodman_08} and probabilistic type systems based on the simply-typed lambda calculus \cite{borgstrom_11}.  Here, we show how a similar approach can be used to provide semantics for a probabilistic type system based on intuitionistic type theory \cite{martinLof_75}, hence including also dependent types.  We call such systems {\em Probabilistic Dependent Type Systems} (PDTS).

We demonstrate that a probabilistic logic can be naturally formulated in our PDTS framework (using the Curry-Howard isomorphism).  The logic we derive is distinct from previous formulations of probablistic logic: Previous approaches have involved associating weights with sentences which induce a distribution over interpretations (possible worlds) \cite{richardson_06}; formulating conditions for consistently assigning probabilities to sentences/interpretations in a simply-typed higher-order language \cite{ng_07,hutter_13}; or assigning terms and deterministic functions probabilistically to types via probabilistic type judgments \cite{cooper_14}.  In contrast, we formulate an intuitionistic type system of non-deterministic functions, and derive a probabilistic logic as a subsytem.  We believe this formulation to be an attractive alternative to the approaches mentioned above, since it arises from a direct application of the Curry-Howard isomorphism.

Sec. \ref{sec:lambdaP} outlines a deterministic dependent type system, which is then generalized to include non-deterministic functions in Sec. \ref{sec:PDTS}, hence forming a PDTS.  Sec. \ref{sec:logic} then uses the PDTS to derive a probabilistic logic via the Curry-Howard isomorphism, and shows that a canonical translation can be made between models formulated using Markov Logic Networks \cite{richardson_06} and PDTS, which are therefore equivalent in expressive power.

\section{A Dependent Type System: $\lambda P^{\Sigma\BOOL}$}\label{sec:lambdaP}

We give an initial overview of a system we will call $\lambda P^{\Sigma\BOOL}$.  This will be a typed lambda calculus with dependent products (hence based on the system $\lambda P$, see \cite{barendregt_92}), dependent sums ($\Sigma$-types), and $\BOOL$ and $\UNIT$ base types.

The language includes the following sets of base constants $\mathcal{C}_{0}$, and sorts, denoted $\mathcal{S}$:
\begin{eqnarray}\label{eq:consts}
\mathcal{C}_0 &::=& 1 \;\;|\;\; \TRUE \;\;|\;\; \FALSE \;\;|\;\;  \UNIT \;\;|\;\; \BOOL \nonumber \\
\mathcal{S} &::=& * \;\;|\;\; \square
\end{eqnarray}
We denote the constants collectively as $\mathcal{C} = \mathcal{C}_{0}\cup\mathcal{S}$. We also have a countably infinite supply of variables, $\mathcal{V} = \{a, b, c, ... A, B, C ... a_1, b_1, c_1 ... a_2 ...\}$.  An abstract syntax (see \cite{barendregt_92}) for {\em values} $\mathcal{U}$ and {\em pseudo-expressions} $\mathcal{T}$ in our language can be specified as follows:
\begin{eqnarray}\label{eq:exp}
\mathcal{U} &::=& \mathcal{C}\;\;|\;\;\mathcal{V}\;\;|\;\;(\mathcal{U},\mathcal{U})_\mathcal{T} \nonumber \\
\mathcal{T} &::=& \mathcal{U} \;\;|\;\; (\mathcal{T} \mathcal{T})  \;\;|\;\; \lambda \mathcal{V}:\mathcal{T}.\mathcal{T} \;\;|\;\;
\IF \mathcal{V} \THEN \mathcal{T} \ELSE \mathcal{T}  \;\;|\;\;  \nonumber \\&&
 \pi_{1} \mathcal{T} \;\;|\;\; \pi_{2} \mathcal{T} \;\;|\;\;
 \sum \mathcal{V}:\mathcal{T}.\mathcal{T} \;\;|\;\; \prod \mathcal{V}:\mathcal{T}.\mathcal{T}
\end{eqnarray}
We define inductively the {\em free variables} of a pseudo-expression $FV(\tau)$, such that $FV(c)=\emptyset$ for $c\in\mathcal{C}$, $FV(v)=\{v\}$ for $v\in\mathcal{V}$, $FV((u_1,u_2))=FV(u_1)\cup FV(u_2)$ for $u_1,u_2\in\mathcal{U}$,  $FV(\tau_1 \tau_2)=FV(\tau_1)\cup FV(\tau_2)$ for $\tau_1,\tau_2\in\mathcal{T}$, $FV(\pi_1 \tau) = FV(\pi_2 \tau) = FV(\tau)$, $FV(\IF v \THEN \tau_1 \ELSE \tau_2)=FV(\tau_1)\cup FV(\tau_2)\cup FV(v)$, and  $FV(\lambda v:\tau_1.\tau_2) = FV(\sum v:\tau_1.\tau_2) = FV(\prod v:\tau_1.\tau_2) = FV(\tau_1)\cup FV(\tau_2)-\{v\}$.  Any variable appearing in $\tau$ which is not a free variable is bound. Pseudo-expressions which can be converted into one another by renaming variables such that no free variable becomes bound after renaming are identified as equivalent (by alpha conversion), and we write $\tau_1 \equiv \tau_2$ when this is the case.  Further, the notion of beta reduction between pseudo-expressions, denoted $\tau_1 \rightarrow_\beta \tau_2$, is specified by the following rules:
\begin{eqnarray}\label{eq:betaReduction}
\pi_1 (u_1,u_2)_\tau &\rightarrow_\beta& u_1 \nonumber \\
\pi_2 (u_1,u_2)_\tau &\rightarrow_\beta& u_2 \nonumber \\
\IF \TRUE \THEN \tau_1 \ELSE \tau_2 &\rightarrow_\beta& \tau_1 \nonumber \\
\IF \FALSE \THEN \tau_1 \ELSE \tau_2 &\rightarrow_\beta& \tau_2 \nonumber \\
(\lambda x:\tau_1.\tau_2) \tau_3 &\rightarrow_\beta& \tau_2[x:=\tau_3] \quad \text{(see conditions below)},
\end{eqnarray}
where $u_1, u_2 \in \mathcal{U}$, $\tau_1,\tau_2,\tau_3, \tau \in \mathcal{T}$, $b_1, b_2 \in \{\TRUE, \FALSE\}$, and $\tau_2[x:=\tau_3]$ denotes the result of substituting $\tau_3$ for $x$ in $\tau_2$.  For rule 5, we also impose the conditions that no variable in $FV(\tau_3)$ becomes bound on substitution (transforming $\tau_2$ by alpha conversion if necessary before substitution to prevent this), and that $\tau_3 \in \mathcal{U}$ if $\tau_2$ contains any sub-expressions of the form $(x,y),(y,x)$ or $(x,x)$.  In addition to the rules above, we extend the beta reduction relation as follows: letting $\tau_2$ be a {\em free sub-(pseudo)-expression} of $\tau_1$, meaning that $\tau_1,\tau_2\in\mathcal{T}$, $\tau_2$ is a substring of $\tau_1$, and no free variable of $\tau_2$ is bound in $\tau_1$ (by a $\lambda,\prod$ or $\sum$ construct), we have that $\tau_1 \rightarrow_\beta \tau_1^\prime$ whenever $\tau_2 \rightarrow_\beta \tau_2^\prime$ by one of the above rules (Eq. \ref{eq:betaReduction}), and $\tau_1^\prime$ is the result of replacing $\tau_2$ for $\tau_2^\prime$ in $\tau_1$. Further, we write $\tau_1 =_\beta \tau_2$ iff $\tau_1 \twoheadrightarrow_\beta \tau_2$ or $\tau_2 \twoheadrightarrow_\beta \tau_1$, where $\twoheadrightarrow_\beta$ is the reflexive transitive closure of $\rightarrow_\beta$.

The pseudo-expressions for which we can infer types will be the expressions proper (legal expressions/terms) of the language.  For this purpose, we define a {\em statement} to be of the form $\tau_0 : \tau_1$ (pronounced $\tau_0$ is of type $\tau_1$).  A {\em context} is a finite linearly ordered set of statements, $\Gamma =  <\tau_{1,0} : \tau_{1,1},\tau_{2,0} : \tau_{2,1},...,\tau_{N,0} : \tau_{N,1} >$, where $\tau_{n,0} \in \mathcal{V}$, $n=1...N$, and for $n_1 \neq n_2$, $\tau_{n_1,0} \neq \tau_{n_2,0}$.  We write $<>$ for the empty context, and $\Gamma, x:\tau$ for the result of appending $x:\tau$ to $\Gamma$.  We can then specify rules for type inference as follows:
\begin{prooftree}
\AxiomC{(axioms):\quad $<> \vdash *:\square, \;\; \BOOL:*, \;\; \UNIT:*, \;\; 1:\UNIT, \;\; \TRUE:\BOOL, \;\; \FALSE:\BOOL,$}
\end{prooftree}
\begin{prooftree}
\AxiomC{$\Gamma \vdash A:s$}
\LeftLabel{(start): \quad}
\RightLabel{$x \not\in \Gamma$}
\UnaryInfC{$\Gamma, x:A \vdash x:A$}
\end{prooftree}
\begin{prooftree}
\AxiomC{$\Gamma \vdash A:B$}
\AxiomC{$\Gamma \vdash C:s$}
\LeftLabel{(weakening): \quad}
\RightLabel{$x \not\in \Gamma$}
\BinaryInfC{$\Gamma, x:C \vdash A:B$}
\end{prooftree}
\begin{prooftree}
\AxiomC{$\Gamma \vdash A:*$}
\AxiomC{$\Gamma, x:A \vdash B:s$}
\LeftLabel{(type/kind formation): \quad}
\BinaryInfC{$\Gamma \vdash (\prod x:A.B):s, \;\; (\sum x:A.B):s$}
\end{prooftree}
\begin{prooftree}
\AxiomC{$\Gamma \vdash F:(\prod x:A.B)$}
\AxiomC{$\Gamma \vdash a:A$}
\LeftLabel{(application): \quad}
\BinaryInfC{$\Gamma \vdash F a:B[x :=a]$}
\end{prooftree}
\begin{prooftree}
\AxiomC{$\Gamma, x:A \vdash b:B$}
\AxiomC{$\Gamma \vdash (\prod x:A.B):s$}
\LeftLabel{(abstraction): \quad}
\BinaryInfC{$\Gamma \vdash (\lambda x:A.b):(\prod x:A.B)$}
\end{prooftree}
\begin{prooftree}
\AxiomC{$\Gamma \vdash a:\BOOL$}
\AxiomC{$\Gamma \vdash b_1:B[x:=\TRUE]$}
\AxiomC{$\Gamma \vdash b_2:B[x:=\FALSE]$}
\LeftLabel{(if): \quad}
\TrinaryInfC{$\Gamma \vdash (\IF a \THEN b_1\ELSE b_2) : B[x:=a]$}
\end{prooftree}
\begin{prooftree}
\AxiomC{$\Gamma, x:A \vdash b:B$}
\AxiomC{$\Gamma \vdash (\sum x:A.B):s$}
\LeftLabel{(products (1)): \quad}
\RightLabel{$x,b \in \mathcal{U}$}
\BinaryInfC{$\Gamma \vdash (x,b)_{(\sum x:A.B)}:(\sum x:A.B)$}
\end{prooftree}
\begin{prooftree}
\AxiomC{$\Gamma \vdash c:(\sum x:A.B)$}
\LeftLabel{(products (2)): \quad}
\UnaryInfC{$\Gamma \vdash (\pi_1 c):A, \;\; (\pi_2 c): (B[x:=(\pi_1 c)])$}
\end{prooftree}
\begin{prooftree}
\AxiomC{$\Gamma \vdash A:B$}
\AxiomC{$\Gamma \vdash C:s$}
\AxiomC{$B =_\beta C$}
\LeftLabel{(type conversion): \quad}
\RightLabel{,}
\TrinaryInfC{$\Gamma \vdash A:C$}
\end{prooftree}
\begin{eqnarray}\label{eq:typeInf}
\end{eqnarray}
where $s$ ranges over $\mathcal{S}$ and $\tau_1 \rightarrow \tau_2$ is shorthand for $\prod x:\tau_1.\tau_2$ in the case that $\tau_2$ is independent of $x$.  In the following, we will also use the notation $\tau_1 \times \tau_2$ for $\sum x:\tau_1.\tau_2$ where $\tau_2$ is independent of $x$, ($\LET x=\tau_1 \IN \tau_2$) for $(\lambda x:A.\tau_2) \tau_1$, where $\tau_1:A$, and $(x,y,z)$ for $(x,(y,z))$ and similarly for longer tuples, where we omit tuple type-tags for clarity.

The system $\lambda P^{\Sigma \BOOL}$ forms a subsystem of the most general system considered in \cite{thompson_91} (TT), and as such can be shown to be strongly normalizing (see Theorem 5.36, \cite{thompson_91}), implying that $\beta$ reduction sequences of all valid expressions can always be extended to terminate in a normal form (an expression which cannot be reduced further), and that each expression has a unique normal form up to alpha equivalence.  We can thus write $\NORM(\tau)$ for the unique normal form of $\tau$, where $\tau$ is a typable expression.  Further properties which can be demonstrated in $\lambda P^{\Sigma\BOOL}$ include {\em preservation} ($\beta$-reduction preserves types up to beta equivalence, in that for a reduction step $\tau_1 \rightarrow_{\beta} \tau_2$, where $\tau_1:\tau_3$ and $\tau_2:\tau_4$, we have that $\tau_3=_{\beta}\tau_4$), and {\em progress} (any expression not in normal form has a valid $\beta$ reduction).  The second of these properties follows directly from strong normalization, and we give a proof of the first in the Appendix (Prop. 2).

\section{Probabilistic Dependent Types: $\lambda P \rho^{\Sigma\BOOL}$}\label{sec:PDTS}

We now enlarge the $\lambda P^{\Sigma\BOOL}$ system to include primitives for sampling from distributions, using a similar syntax to \cite{borgstrom_11}.  The calculus developed in \cite{borgstrom_11} is based on the simple typed lambda calculus, and includes only distributions over values within a type.  Since our calculus also includes dependent types, we naturally derive expressions which represent distributions over terms of multiple types, distributions over types, as well as functions whose return type is probabilistically determined.  The language we develop in this section will be called $\lambda P \rho^{\Sigma \BOOL}$, and contains as a subsystem the fragment of the Fun language in \cite{borgstrom_11} only involving Bernoulli distributions (and without $\OBSERVE$ statements)\footnote{The $\OBSERVE$ statement in \cite{borgstrom_11} allows a distribution to be conditioned on (and updated by) evidence.  As we discuss in Sec. \ref{subsec:MLN}, such conditioning can be simulated by adding the conditioning formula with a weight tending to $\infty$ in the context of Markov Logic Networks, and hence also in Dependent Type Networks by the correspondence we describe in Sec. \ref{subsec:proof}.  Further, in a general PDTS we could say return the unit value from a probabilistic function if the conditional is not satisfied.  We thus simplify our analysis by not including the statement in our syntax.} when the dependent types are removed.

We begin by expanding the syntax of Sec. \ref{sec:lambdaP}:
\begin{eqnarray}\label{eq:exp2}
\mathcal{U} &::=& \mathcal{C}_{0}\;\;|\;\;\mathcal{V}\;\;|\;\;(\mathcal{U},\mathcal{U})_\mathcal{T} \nonumber \\
\mathcal{Z} &::=& \{\} \;\;|\;\; \{(\mathcal{T},\mathcal{T}),\;\;\mathcal{Z}\} \nonumber \\
\mathcal{T} &::=& \mathcal{U} \;\;|\;\; (\mathcal{T} \mathcal{T})  \;\;|\;\; \lambda \mathcal{V}:\mathcal{T}.\mathcal{T} \;\;|\;\;
 \IF \mathcal{V} \THEN \mathcal{T} \ELSE \mathcal{T}  \;\;|\;\;  \nonumber \\&&
 \pi_{1} \mathcal{T} \;\;|\;\; \pi_{2} \mathcal{T} \;\;|\;\;
 \sum \mathcal{V}:\mathcal{T}.\mathcal{T} \;\;|\;\; \prod \mathcal{V}:\mathcal{T}.\mathcal{T}  \;\;|\;\;  (\lambda \mathcal{V}.\mathcal{Z}) \mathcal{T} \;\;|\;\; \RANDOM_\rho(\mathcal{T})
\end{eqnarray}
where $\rho$ is any real number in the interval $(0\;1)$, and the new syntactic construction $\mathcal{Z}$ allows us to form lists/ordered sets of paired terms.  Since the syntax in Eq. \ref{eq:exp2} is an expansion of that in Eq. \ref{eq:exp}, all pseudo-expressions of $\lambda P^{\Sigma\BOOL}$ will also be pseudo-expressions of $\lambda P \rho^{\Sigma\BOOL}$.  We denote the set of common pseudo-expressions as $\mathcal{T}_{\lambda P^{\Sigma\BOOL}}$.  Next, we define inductively a weighted reduction relationship ($\beta\rho$-reduction) on pseudo-expressions, $\tau_1 \rightarrow^{\rho}_{\beta} \tau_2$ via:
\begin{eqnarray}\label{eq:betaReduction2}
\pi_1 (u_1,u_2)_\mathcal{T} &\rightarrow^1_\beta& u_1 \nonumber \\
\pi_2 (u_1,u_2)_\mathcal{T} &\rightarrow^1_\beta& u_2 \nonumber \\
\IF \TRUE \THEN \tau_1 \ELSE \tau_2 &\rightarrow^1_\beta& \tau_1 \nonumber \\
\IF \FALSE \THEN \tau_1 \ELSE \tau_2 &\rightarrow^1_\beta& \tau_2 \nonumber \\
\RANDOM_{\rho}(\tau) &\rightarrow^{\rho}_\beta& (\tau \TRUE) \nonumber \\
\RANDOM_{\rho}(\tau) &\rightarrow^{(1-\rho)}_\beta& (\tau \FALSE) \nonumber \\
(\lambda x:\tau_1.\tau_2) \tau_3 &\rightarrow^1_\beta& \tau_2[x:=\tau_3] \quad \text{if $\tau_3\in\mathcal{T}_{\lambda P^{\Sigma\BOOL}}$, and conditions as in Eq. \ref{eq:betaReduction} rule 5,}\nonumber \\
(\lambda x.Z) \tau_3 &\rightarrow^1_\beta& (\lambda x:\tau_1.\tau_2)\tau_3 \quad \text{if $\tau_3\in\mathcal{T}_{\lambda P^{\Sigma\BOOL}}$, $\tau_3:\tau_1$, $(\tau_1,\tau_2)\in\in Z$,} \nonumber \\
 && \quad\quad\quad \text{and $\forall \tau_2^\prime \neq \tau_2, \neg((\tau_1,\tau_2^\prime) \in\in Z)$.}
\end{eqnarray}
where, as before, $u_1, u_2 \in \mathcal{U}$, $\tau_1,\tau_2,\tau_3, \tau \in \mathcal{T}$, $b_1, b_2 \in \{\TRUE, \FALSE\}$, and here $Z\in \mathcal{Z}$, while we write $\in\in$ for the transitive closure of the set membership relation where $Z$ is interpreted as denoting a set (alternatively, $\in\in$ denotes list membership if $Z$ is viewed as a list data structure).  We extend the reduction relation by allowing $\tau_1 \rightarrow^{\rho}_\beta \tau_1^\prime$ whenever $\tau_2 \rightarrow^{\rho}_\beta \tau_2^\prime$ by one of the above rules, and $\tau_1^\prime$ is the result of replacing $\tau_2$ for $\tau_2^\prime$ in $\tau_1$, where $\tau_2$ is either (i) the leftmost-outermost reducible sub-(psuedo)-expression of the form $\tau_2=\RANDOM_{\rho}(\tau_3)$ or $\tau_2=(\lambda x.Z)\tau_3$, or (ii) if no reducible sub-(psuedo)-expression of the forms in (i) exist, the leftmost-outermost free sub-(pseudo)-expression of $\tau_1$ as in Sec. \ref{sec:lambdaP}.  Here, {\em leftmost-outermost} refers to the sub-expression found by searching the parse tree of a (pseudo)-expression top-down, traversing the left subtree of a non-reducible expression before the right (see \cite{thompson_91}, Def. 2.13).  These conditions ensure that there is at most one sub-(pseudo)-expression whose reduction is allowed at any point.  The relationship $\tau_1 \rightarrow^{\rho}_{\beta} \tau_2$ can be given the operational semantics `$\tau_1$ reduces to $\tau_2$ with probability $\rho$'.  We can further define the relation $\tau_1 \twoheadrightarrow_\beta \tau_2$ to hold when there exists any reduction chain between $\tau_1$ and $\tau_2$ such that no step has zero probability, and $\tau_1 \twoheadrightarrow^{\rho_0}_\beta \tau_2$ to hold when the sum of the products of the $\rho$ values across all such chains is $\rho_0$ (where $\rho_0\leq 1$, see Appendix Prop. 5).  Further, we note that the rules for reduction of $\lambda$ terms in Eq. \ref{eq:betaReduction2} whereby a term may not be substituted until it is reduced to a term in $\mathcal{T}_{\lambda P^{\Sigma\BOOL}}$ enforce a `stochastic memoization' property on reduction sequences as defined in \cite{goodman_08} (a $\RANDOM$ statement will not be duplicated by substitution, and reduced in alternative ways in a single reduction sequence).

In \cite{borgstrom_11}, type assignment rules are given which consistently assign a single type to all expressions such that a {\em preservation} property holds: in any probabilistic reduction sequence, the types of all terms are preserved at each reduction step (up to $\beta$ equivalence).  Since our language contains dependent types, such a property cannot be expected to hold, as a given expression may reduce to expressions with different types by different reduction sequences.  Instead of assigning types to all expressions in $\lambda P \rho^{\Sigma\BOOL}$, only expressions already typed in $\lambda P^{\Sigma\BOOL}$ by the rules of Eq. \ref{eq:typeInf} will be typed directly.  To identify general valid expressions in $\lambda P \rho^{\Sigma\BOOL}$, we define two operators $\TYPES$ and $\REDUCTIONS$ on pseudo-expressions, where $\REDUCTIONS(\tau)$ will return the set of normal forms to which a given psuedo-expression reduces with non-zero probability, and $\TYPES(\tau)$ will return the set of types corresponding to those normal forms (we will ensure all reduction sequences of legal expressions in $\lambda P \rho^{\Sigma\BOOL}$ end in $\lambda P^{\Sigma\BOOL}$ normal forms).  We define these operators inductively.  For each pseudo-expression $\tau_1$ of $\lambda P^{\Sigma\BOOL}$, we set $\TYPES(\tau_1)=\{\tau_2\}$ when $\tau_1:\tau_2$, $\tau_2\in\NF_{\lambda P^{\Sigma\BOOL}}$, and $\TYPES(\tau_1)=\{\NOTYPE\}$ otherwise.  Further, $\REDUCTIONS(\tau_1)=\{\NORM(\tau_1)\}$ in the former case, and $\emptyset$ in the latter.  The following inductive rules are then used to define these operators for all valid expressions in $\lambda P \rho^{\Sigma\BOOL}$ (where  we let $\mathcal{T}_x$ be the set of $\lambda P \rho^{\Sigma\BOOL}$ pseudo-expressions containing no reducible sub-expressions containing the variable $x$, including sub-expressions within $\mathcal{Z}$ constructions):
\begin{prooftree}
\AxiomC{$\Gamma \vdash \REDUCTIONS(F)=\{\lambda x:\BOOL.a_n, n=1...N_1\},$}
\noLine
\UnaryInfC{$\Gamma \vdash \TYPES(F)=\{\prod x:\BOOL.A_n, n=1...N_2\},$}
\LeftLabel{(random): \quad}
\UnaryInfC{$\Gamma \vdash \REDUCTIONS(\RANDOM_\rho(F)) = \bigcup_{b \in \{\TRUE,\FALSE\}}\{\NORM(a_n[x:=b]), n = 1...N_1\}$}
\noLine
\UnaryInfC{$\Gamma \vdash \TYPES(\RANDOM_\rho(F)) = \bigcup_{b \in \{\TRUE,\FALSE\}}\{\NORM(A_n[x:=b]), n = 1...N_2\}$}
\end{prooftree}
\begin{prooftree}
\AxiomC{$\Gamma \vdash a:\BOOL, \REDUCTIONS(a_1)=\{b_{1,n}, n = 1...N_{1,1}\},$}
\noLine
\UnaryInfC{$\REDUCTIONS(a_2)=\{b_{2,n}, n = 1...N_{2,1}\},$}
\noLine
\UnaryInfC{$\TYPES(a_1)=\{B_{1,n}, n = 1...N_{1,2}\},$}
\noLine
\UnaryInfC{$\TYPES(a_2)=\{B_{2,n}, n = 1...N_{2,2}\}$}
\LeftLabel{(if): \quad}
\UnaryInfC{$\Gamma \vdash \REDUCTIONS(\IF a \THEN a_1 \ELSE a_2) = \{\IF a \THEN b_{1,n_1} \ELSE b_{2,n_2}, n_1 = 1...N_{1,1}, n_2 = 1...N_{2,1}\}$}
\noLine
\UnaryInfC{$\Gamma \vdash \TYPES(\IF a \THEN a_1 \ELSE a_2) = \{\IF a \THEN B_{1,n_1} \ELSE B_{2,n_2}, n_1 = 1...N_{1,2}, n_2 = 1...N_{2,2}\}$}
\end{prooftree}
\begin{prooftree}
\AxiomC{$\Gamma,  x:A \vdash \REDUCTIONS(b)=\{b_n, n = 1...N_1\},$}
\noLine
\UnaryInfC{$\Gamma,  x:A \vdash \TYPES(b)=\{B_n, n = 1...N_2\}, \quad b \in \mathcal{T}_x$}
\LeftLabel{(abstraction): \quad}
\UnaryInfC{$\Gamma \vdash \REDUCTIONS(\lambda x:A.b) = \{\lambda x:A.b_n, n = 1...N_1\}$}
\noLine
\UnaryInfC{$\Gamma \vdash \TYPES(\lambda x:A.b) = \{\prod x:A.B_n, n = 1...N_2\}$}
\end{prooftree}
\begin{prooftree}
\AxiomC{$\Gamma \vdash \REDUCTIONS(a)=\mathcal{A}, \REDUCTIONS(F)=\{\lambda x:A.b_n, n=1...N_1\},$}
\noLine
\UnaryInfC{$\Gamma \vdash \TYPES(a)=\{A\}, \TYPES(F)=\{\prod x:A.B_n, n=1...N_2\}$}
\LeftLabel{(application (1)): \quad}
\UnaryInfC{$\Gamma \vdash \REDUCTIONS(F a) = \bigcup_{a^\prime \in \mathcal{A}}\{\NORM(b_n[x:=a^\prime]), n = 1...N_1\}$}
\noLine
\UnaryInfC{$\Gamma \vdash \TYPES(F a) = \bigcup_{a^\prime \in \mathcal{A}}\{\NORM(B_n[x:=a^\prime]), n = 1...N_2\}$}
\end{prooftree}
\begin{prooftree}
\AxiomC{$\Gamma \vdash \REDUCTIONS(a)=\mathcal{A}, \TYPES(a)=\mathcal{A}^\prime$,}
\noLine
\UnaryInfC{$\forall A \in \mathcal{A}^\prime, (A:*)\wedge (\exists (A,b) \in\in Z, \st \forall b^\prime \neq b, \neg((A,b^\prime)\in\in Z)\wedge$}
\noLine
\UnaryInfC{$(\forall a^\prime \in \mathcal{A} \st a^\prime:A, \forall B\in\TYPES(b[x:=a^\prime]),B:s)\wedge(b\in\mathcal{T}_x))$}
\LeftLabel{(application (2)): \quad}
\UnaryInfC{$\Gamma \vdash \REDUCTIONS((\lambda x.Z) a) = \bigcup_{a^\prime \in \mathcal{A}, (A,b)\in\in Z, a^\prime:A}\REDUCTIONS(b[x:=a^\prime])$}
\noLine
\UnaryInfC{$\Gamma \vdash \TYPES((\lambda x.Z) a) = \bigcup_{a^\prime \in \mathcal{A}, (A,b)\in\in Z, a^\prime:A}\TYPES(b[x:=a^\prime])$}
\end{prooftree}
\begin{prooftree}
\AxiomC{$\Gamma \vdash A:*, \quad \Gamma,  x:A \vdash \REDUCTIONS(B)=\{B_n, n = 1...N_1\}, $}
\noLine
\UnaryInfC{$\Gamma,  a:A \vdash \TYPES(B)=\{s\}, \quad s \in \{*,\square\}, \quad B \in \mathcal{T}_x$}
\LeftLabel{($\prod$/$\sum$ formation): \quad}
\UnaryInfC{$\Gamma \vdash \REDUCTIONS(\prod x:A.B) = \{\prod x:A.B_n, n = 1...N_1\}$}
\noLine
\UnaryInfC{$\Gamma \vdash \REDUCTIONS(\sum x:A.B) = \{\sum x:A.B_n, n = 1...N_1\}$}
\noLine
\UnaryInfC{$\Gamma \vdash \TYPES(\prod x:A.B) = \{s\}$}
\noLine
\UnaryInfC{$\Gamma \vdash \TYPES(\sum x:A.B) = \{s\}$}
\end{prooftree}
\begin{prooftree}
\AxiomC{$\Gamma \vdash \REDUCTIONS(a)=\mathcal{A}, \TYPES(a)=\{\prod a:A.B\}$}
\LeftLabel{(products): \quad}
\UnaryInfC{$\Gamma \vdash \REDUCTIONS(\pi_1 a) = \bigcup_{a^\prime \in \mathcal{A}}\{\NORM(\pi_1 a^\prime)\}$}
\noLine
\UnaryInfC{$\Gamma \vdash \REDUCTIONS(\pi_2 a) = \bigcup_{a^\prime \in \mathcal{A}}\{\NORM(\pi_2 a^\prime)\}$}
\noLine
\UnaryInfC{$\Gamma \vdash \TYPES(\pi_1 a) = \{A\}$}
\noLine
\UnaryInfC{$\Gamma \vdash \TYPES(\pi_2 a) = \bigcup_{a^\prime \in \mathcal{A}}\{\NORM(B[x:=\pi_1 a^\prime])\}$.}
\end{prooftree}
\begin{eqnarray}\label{eq:typeInf2}
\end{eqnarray}
All pseudo-expressions $\tau$ whose values cannot be set by Eq. \ref{eq:typeInf2} are set to $\TYPES(\tau) = \{\NOTYPE\}$, $\REDUCTIONS(\tau) = \emptyset$.  The expressions proper (legal expressions) of $\lambda P \rho^{\Sigma\BOOL}$ are all those for which $\NOTYPE \not\in \TYPES(\tau)$ and $\exists s\in\mathcal{S} \st \forall \tau^\prime \in \TYPES(\tau), \tau^\prime:s$ (an expression can reduce to either types/type-constructors or terms belonging to types, not both).

As a consequence of Eqs. \ref{eq:betaReduction2} and \ref{eq:typeInf2}, we can show that for any legal expression $\tau_1$ and reduction step $\tau_1 \rightarrow^{\rho}_{\beta} \tau_2$, we have that $\TYPES(\tau_1)\supseteq\TYPES(\tau_2)$, and $\REDUCTIONS(\tau_1)\supseteq\REDUCTIONS(\tau_2)$ (Appendix, Prop. 3).  Hence, in place of a {\em strong preservation} property in the framework of \cite{borgstrom_11} in which probabilistic reduction preserves types, we have a {\em weak preservation} property in which probabilistic reduction preserves a non-empty subset of types.  Further, we can show that a {\em progress} property holds for $\rightarrow^\rho_\beta$ reduction, in that for any pseudo-expression $\tau_1$ not in $\lambda P^{\Sigma\BOOL}$ normal form there exists a $\rho$ and $\tau_2$ such that $\tau_1 \rightarrow^\rho_\beta \tau_2$, and that $\sum_{\{(\rho,\tau_2)|\tau_1 \rightarrow^\rho_\beta \tau_2\}}\rho = 1$ (i.e. we can make a probabilistic reduction step, see Appendix, Prop. 4a), and that all valid reduction sequences of expressions in $\lambda P \rho^{\Sigma\BOOL}$ end in an expression of $\lambda P^{\Sigma\BOOL}$ in normal form (Appendix, Prop. 4b).  Finally, we can show that for any $\tau_3 \in \REDUCTIONS(\tau_1)$, $\tau_1 \twoheadrightarrow^\rho_\beta \tau_3$ with $\rho > 0$, and (a) for each $\tau_3 \in \REDUCTIONS(\tau_1)$, $\exists \tau_4 \in \TYPES(\tau_1) \st \tau_3:\tau_4$; (b) for each $\tau_4 \in \TYPES(\tau_1)$, $\exists \tau_3 \in \REDUCTIONS(\tau_1) \st \tau_3:\tau_4$, i.e. $\TYPES(\tau_1)$ contains exactly the normal form types resulting from reductions of $\tau_1$ for which $\rho>0$ (Appendix, Prop. 6).

Taken together, the above properties imply that {\em probabilistic type judgement} is possible by sampling in $\lambda P \rho^{\Sigma\BOOL}$.  Letting $\iota_{\tau_1}(\tau_2)=\sum_{\{\tau_3\in\NF_{\lambda P^{\Sigma\BOOL}}|\tau_3:\tau_2\}}\rho_{\tau_3}$, where $\tau_1 \twoheadrightarrow^{\rho_{\tau_3}}_{\beta} \tau_3$, we can think of $\iota_{\tau_1}(.)$ as a characteristic function for $\TYPES(\tau_1)$, where $\iota_{\tau_1}(\tau_2)>0\Rightarrow \tau_2\in\TYPES(\tau_1)$ and $\iota_{\tau_1}(\tau_2)=0\Rightarrow \tau_2\not\in\TYPES(\tau_1)$.  We can estimate $\iota_{\tau_1}(\tau_2)$ by repeated $\beta\rho$ reduction of $\tau_1$ to $\tau_3$, taking the frequency with which $\tau_3:\tau_2$ by using the fact that derivability of type judgments is decidable  in $\lambda P^{\Sigma\BOOL}$ (Appendix, Prop. 7).

Expressions in the system $\lambda P \rho^{\Sigma\BOOL}$ can be given a two-level semantics in terms of probability distributions.  Most directly, expressions can place a low level distribution across terms which they reduce to of several types, which induces a higher level distribution across the types themselves ($\iota_{\tau_1}(.)$ above).  Alternatively, since expressions in $\lambda P^{\Sigma\BOOL}$ may also reduce to types and type-constructors belonging to {\em kinds} such as $*$ or $A \rightarrow *$, they may induce a low level distribution across the former entities, and a higher level distribution across kinds.  In each case, reduction of an expression corresponds to sampling explicitly from the low level distribution and implicitly from the high level distribution (a {\em sampling semantics}).  We briefly note some special cases here (assuming an empty context).  If $\TYPES(\tau)=\{A\}$, $A:*$, $\tau$ represents a (non-dependent) distribution over a single type.  When $A:\square$, $\tau$ is a distribution over a kind, for instance a distribution over types themselves ($A=*$), or type-constructors ($A=B \rightarrow *$).  If $\TYPES(\tau)=\{A,B,C\}$, $A,B,C:*$, then $\tau$ is distribution over terms of types $A,B$ and $C$ at the low level, and a distribution over the types at the higher level.  If in addition $\tau$ has the form $F x$, where $\TYPES(x)=\{X\}$, then $\tau$ can be thought of as a mixture distribution, where the components are determined by the distribution over $X$ denoted by $x$.  Alternatively, for $\TYPES(\tau)=\{\prod x:A.B_n, n=1...N_1\}$, and $B_n[x:=a]:*$, $\tau$ can be thought of both as a distribution across functions with types in $\TYPES(\tau)$, and as a function itself, which when given a distribution $\tau^\prime$ across $A$ returns a distribution across the return types of the functions in $\TYPES(\tau)$.  We note also how expressions of the form $(\lambda x.Z) \tau$ can be used to transform a general distribution across terms of several types represented by $\tau$ into another general distribution across terms of several types.

\section{Relationship to Markov Logic Networks via Formulae as Types Interpretation}\label{sec:logic}

We now discuss the relationship between a particular class of PDTS, which we shall call {\em Dependent Type Networks} (DTNs), and Markov Logic Networks (MLNs) as introduced in \cite{richardson_06}.  As we show, these systems are equivalent under a probabilistic analogue of the formulae as types interpretation (or Curry-Howard isomorphism) for typed lambda systems.  The equivalence provides a further semantics for this class of PDTS in terms of MLNs, which in turn can be given factor graph semantics.  Further, it is suggestive of algorithmic possibilities in that algorithms for performing inference in either of the representations can be transformed to perform inference in the other.

\subsection{Markov Logic Networks}\label{subsec:MLN}

As defined in \cite{richardson_06}, a Markov Logic Network $L$ is a set of pairs $(F_i,w_i)$, where $F_i$ is a formula in first order logic (constructed using constants, variables, functions, predicates and logical connectives) and $w_i$ is a real number.  $L$ acts as a template, which when combined with a finite set of constants $C = \{c_1,c_2,...,c_{|C|}\}$ (which includes and may extend the constants used to build the $F_i$'s, defining the domain over which the model ranges) constructs a {\em ground Markov Network}, $M_{L,C}$.  $M_{L,C}$ contains a binary node for every possible grounding of the predicates in $L$ (a {\em ground predicate} being a formula involving only the applications of a predicate to a tuple of constants).  For instance, if $P_1$ is a unary predicate, $M_{L,C}$ contains a node for each application $P_1(c_i), i=1...|C|$, and if $P_2$ is a binary predicate, a node for each application $P_2(c_i,c_j), i,j=1...|C|$.  The network also contains cliques for all possible groundings of each formula $F$ in $L$ (i.e. substitutions of constants for free variables in $F$)\footnote{Formulas of the form $\forall x F$ are grounded as a conjunction of all possible groundings of $x$ in $F$, and $\exists x F$ as a disjunction of all possible groundings.  We will not assume that $L$ is automatically transformed into conjunctive normal form, and so distinguish between $L=\{(F_1,0.5*w),(F_2,0.5*w)\}$ and $L=\{(F_1\wedge F_2,w)\}$, as well as $L=\{(\forall x F,|C|w)\}$ and $L=\{(F,w)\}$ ($FV(F)=\{x\}$), where the latter can be taken as shorthand for $L=\{(F[x:=c_1],w),(F[x:=c_2],w)...(F[x:=c_{|C|}],w)\}$ (which is only equivalent to $L=\{(\forall x F,|C|w)\}$ if transformation to conjunctive normal form is assumed).}, where the clique derived from the $j$'th grounding of formula $i$ is associated with a clique potential $\psi_{i,j}$ which is $1$ if the ground formula is not satisfied, and $e^{w_i}$ if it is.  A {\em possible world} is fixed by a joint setting of the nodes of $M_{L,C}$ (under the assumptions of unique names, domain closure and known functions, see \cite{richardson_06}), which we denote by $\mathbf{x}\in\{0,1\}^P$, where $P$ is the number of ground predicates (nodes) in the network.  $M_{L,C}$ thus defines a distribution over possible worlds, which can be expressed as a random variable $X$ which ranges over the settings $\mathbf{x}$:
\begin{eqnarray}\label{eq:MLN}
P(X=\mathbf{x}) = \frac{1}{Z}\prod_{ij} \psi_{ij}(\mathbf{x}_{ij}) = \frac{1}{Z}\prod_{i} (e^{w_i})^{n_i(\mathbf{x})},
\end{eqnarray}
where $Z$ is the partition function $Z = \sum_{\mathbf{x}} \prod_{ij} \psi_{ij}(\mathbf{x}_{ij})$, and $n_i(\mathbf{x})$ denotes the number of true groundings of formula $F_i$ in joint setting $\mathbf{x}$.  We note that the distribution defined by Eq. \ref{eq:MLN} is maximum entropy in the sense that any ground predicates which are independent of the weighted $F_i$'s (hence {\em probabilistically undecidable}) will be true or false with equal probability.

For convenience, we consider only unconditioned inference.  Here, we are interested in the probability that a given formula holds in a possible world distributed according to Eq. \ref{eq:MLN}, which can be written $P(F|L,C) = \sum_{\mathbf{x}\in\mathcal{X}_F}P(\mathbf{x}|L,C)$, where $\mathcal{X}_F$ is the subset of possible worlds in which query formula $F$ holds (conditional queries can be simulated by adding the conditioning formula $F^\prime$ to the MLN with a weight tending to $\infty$).   We give two examples below.  First, consider an MLN $L$ with a formula $F_1$ corresponding to $A(x)\rightarrow B(x)$ (equivalently $B(x)\vee\neg A(x)$) with weight $w_1$, where $A,B$ are unary predicates and $\rightarrow$ implication, and $F_2 = A(c_1)$ with weight $w_2$ and $c_1$ a constant ($w_1,w_2 > 0$).  We then consider the grounded network over over $C=\{c_1\}$, and the query $B(c_1)$.  Writing $[b_1 b_2]$ for the setting of $\mathbf{x}$ which implies $A(c_1)$ when $b_1=1$ and $\neg A(c_1)$ when $b_1=0$ and similarly for $B(c_1)$ and $b_2$, the query has the probability:
\begin{eqnarray}\label{eq:MLNex1}
P(B(c_1)|L,C) = P(\mathbf{x}=[0\;\;1]) + P(\mathbf{x}=[1\;\;1]) = \frac{e^{w_1}(1+e^{w_2})}{e^{w_1}(2+e^{w_2})+e^{w_2}},
\end{eqnarray}
which tends to $1$ as $w_1,w_2$ tend to $\infty$.  As a second example, consider an MLN $L$ with a formula $F_1=A(x)$, with weight $w_1>0$ and $A(x)$ a unary predicate.  We then consider the grounded network over $C=\{c_1,c_2\}$, and the query $\exists x A(x)$ (equivalently $A(c_1)\vee A(c_2)$, under the assumptions above).  Writing $[b_1 b_2]$ for the setting of $\mathbf{x}$ which implies $A(c_1)$ when $b_1=1$ and $\neg A(c_1)$ when $b_1=0$ and similarly for $A(c_2)$ and $b_2$, the query has the probability:
\begin{eqnarray}\label{eq:MLNex2}
P(\exists x A(x)|L,C) &=& P(\mathbf{x}=[1\;\;0]) + P(\mathbf{x}=[0\;\;1]) + P(\mathbf{x}=[1\;\;1]) \nonumber \\
&=& \frac{e^{w_1}(2+e^{w_1})}{e^{w_1}(2+e^{w_1})+1},
\end{eqnarray}
which again tends to $1$ as $w_1$ tends to $\infty$.  As detailed in \cite{richardson_06}, queries in larger networks can be estimated by building a factor graph representation of the ground Markov Network, and using algorithms such as Gibbs sampling, Loopy Belief Propagation and variational methods such as mean-field message passing.

\subsection{Dependent Type Networks}\label{subsec:DTN}

We now define the notion of a {\em Dependent Type Network} (DTN), which will be a probabilistic dependent type system in the sense of Sec. \ref{sec:PDTS}.  A `language' for the PDTS will be fixed by specifying a context of the form (assuming only unary and binary predicates/functions):
\begin{eqnarray}\label{eq:PDTcontext}
\Gamma_0 &=& \Gamma^\prime, \Gamma^{\prime\prime} \nonumber \\
\Gamma^\prime &=& <A:*, \bot:*, c_{1...N_C}:A, B^1_{1...N_{B^1}}:A\rightarrow*, \nonumber \\
&& B^2_{1...N_{B^2}}:A\times A\rightarrow*, g^1_{1...N_{g^1}}:A\rightarrow A, g^2_{1...N_{g^2}}:A\times A\rightarrow A> \nonumber \\
\Gamma^{\prime\prime} &=& <b^1_{n=1...N_{B^1},m=1...N_C,1}:B^1_n(c_m),b^1_{n=1...N_{B^1},m=1...N_C,0}:B^1_n(c_m)\rightarrow\bot,\nonumber \\
&& b^2_{n=1...N_{B^2},m=1...N_C,l=1...N_C,1}:B^2_n(c_m,c_l),\nonumber \\
&& b^2_{n=1...N_{B^2},m=1...N_C,l=1...N_C,0}:B^2_n(c_m,c_l)\rightarrow\bot>.
\end{eqnarray}
Here, $A$ specifies a domain, to which constants $c_{1...N_c}$ belong.  $B^1_{n=1...N_{B^1}}$, and $B^2_{n=1...N_{B^2}}$ will be used as unary and binary predicates over the domain respectively, $g^1_{n=1...N_{g^1}}$ and $g^2_{n=1...N_{g^2}}$ as unary and binary functions, $\bot$ to represent a contradiction, and the sets of terms $b^1$ and $b^2$ as ground predicates, while we write $B(c)$ as alternative notation for application.  A DTN is fixed by augmenting this context with a further set of `formulas' in the language, specified by $\Gamma_1 = \Gamma_0 , <f_1:F_1,...,f_{N_F}:F_{N_F}>$.  Here, a given formula such as $B^1(c_1)\wedge B^1(c_2)$ is represented by the type corresponding to the formula by the formulae as types interpretation of typed lambda systems (equivalently, the {\em propositions as types} interpretation, see \cite{barendregt_92}), $B^1(c_1)\times B^1(c_2)$.  In general, $X\wedge Y$ corresponds to $X\times Y$ ($X$ and $Y$ being ground predicates), $X\vee Y$ to the {\em disjoint product} $\sum x:\BOOL.(\IF x \THEN X \ELSE Y)$, $X\rightarrow Y$ to $X\rightarrow Y$, $\neg X$ to $X\rightarrow\bot$, $\forall x (X(x))$ to $\prod x:A.X(x)$, and $\exists x (X(x))$ to $\sum x:A.X(x)$.  Quantifications over binary and higher order predicates can be represented by nested dependent sums and products.  An expression which can be typed to a formula type corresponds to a proof of that formula, and thus to asserting the formula itself (a formula is `true' if it is inhabited).  We will further equip a PDT with a set of probabilistic weights, $P = \{p_1,p_2,...,p_{N_F}\}$, where $p_i$ will represent the probability that $F_i$ is true as a constant in the meta-language, taking values in $(0\;\;1)$.  We summarize the above in the following definition:

\vspace{0.3cm}
\textbf{Definition 1.} \textit{A {\em Dependent Type Network} is a probabilistic dependent type system equipped with a context $\Gamma_0$ representing a language containing a domain, constants, predicates, functions, contradiction and ground predicate symbols as above, a context $\Gamma_1$ which augments $\Gamma_0$ with a set of with a set of formula proofs represented as constants assigned to types in the language ($f_i:F_i, i=1...N_F$), and a set $P$ of meta-language constants $p_i$, which assign a probabilistic weight in $(0\;\;1)$ to each formula.}

\vspace{0.3cm}
Given a DTN as defined, we will represent a query formula, for instance $Q=B^1_1(c_3)$, as an expression (/program) $q_Q$ where $\TYPES(q_Q)=\{B^1_1(c_3),B^1_1(c_3)\rightarrow\bot,\bot\}$.  That is, the expression reduces to either a proof of $Q$, a proof of $\neg Q$, or a proof of a contradiction (the last possibility corresponding to an inconsistent world).  We will canonically define $q_Q$ for a particular $Q$ (where $Q$ is a type) as follows:
\begin{eqnarray}\label{eq:PDTquery}
q_Q &=& \LET x_1 = \RANDOM_{p_1}(\lambda x:\BOOL.\IF x \THEN f_1 \ELSE 1) \IN \nonumber \\
&& ... \nonumber \\
&& \LET x_{N_F} = \RANDOM_{p_{N_F}}(\lambda x:\BOOL.\IF x \THEN f_{N_F} \ELSE 1) \IN \nonumber \\
&& \LET x_{N_F+1} = D \IN \nonumber \\
&& (\lambda x.Z_Q) (x_1,x_2,...,x_{N_F},x_{N_F+1})
\end{eqnarray}
In Eq. \ref{eq:PDTquery}, $D$ is defined as follows:
\begin{eqnarray}\label{eq:PDTquery2}
D &=& \LET x_{1,1} = \RANDOM_{0.5}(\lambda x:\BOOL.\IF x \THEN b_{1,1,1} \ELSE b_{1,1,0}) \IN \nonumber \\
&& ... \nonumber \\
&& \LET x_{N_{B^1},N_C} = \RANDOM_{0.5}(\lambda x:\BOOL.\IF x \THEN b_{N_{B^1},N_C,1} \ELSE b_{N_{B^1},N_C,0}) \IN \nonumber \\
&& \LET x_{1,1,1} = \RANDOM_{0.5}(\lambda x:\BOOL.\IF x \THEN b_{1,1,1,1} \ELSE b_{1,1,1,0}) \IN \nonumber \\
&& ... \nonumber \\
&& \LET x_{N_{B^1},N_C,N_C} = \RANDOM_{0.5}(\lambda x:\BOOL.\IF x \THEN b_{N_{B^2},N_C,N_C,1} \ELSE b_{N_{B^2},N_C,N_C,0}) \IN \nonumber \\
&& (x_{1,1},...,x_{N_{B^1},N_C},x_{1,1,1},...,x_{N_{B^2},N_C,N_C})
\end{eqnarray}
representing an expression which randomly samples a possible world.  The term $Z_Q$ in Eq. \ref{eq:PDTquery} is constructed as a list of pairs $[(t,\tau_t),t\in \TYPES(T)]$, where $T$ is:
\begin{eqnarray}\label{eq:PDTquery3}
T &=& \LET x_1 = \RANDOM_{p_1}(\lambda x:\BOOL.\IF x \THEN f_1 \ELSE 1) \IN \nonumber \\
&& ... \nonumber \\
&& \LET x_{N_F} = \RANDOM_{p_{N_F}}(\lambda x:\BOOL.\IF x \THEN f_{N_F} \ELSE 1) \IN \nonumber \\
&& \LET x_{N_F+1} = D \IN \nonumber \\
&& (x_1,x_2,...,x_{N_F},x_{N_F+1})
\end{eqnarray}
For each $t\in \TYPES(T)$, we consider the context $\Gamma_t = \Gamma^\prime,x:t$, and construct the sets $R_{Q,t}$ of expressions $r$ such that $\Gamma_t \vdash r:Q$, $S_{Q,t}$ of expressions $s$ such that $\Gamma_t \vdash s:Q\rightarrow\bot$, and $K_{Q,t}$ of expressions $k$ such that $\Gamma_t \vdash k:\bot$.  If $K\neq\emptyset$, we set $\tau_t=k$ for an arbitrary $k\in K$.  Otherwise, we must have either $R=\emptyset$ or $S=\emptyset$ (since if neither is the case $(s r)\in K$ for arbitrary $s\in S$, $r\in R$, and the term $D$ in Eq. \ref{eq:PDTquery3} ensures at least one is non-empty).  If $R=\emptyset\wedge K=\emptyset$, we set $\tau_t=s$ for an arbitrary $s\in S$, and if $S=\emptyset\wedge K=\emptyset$, we set $\tau_t=r$ for an arbitrary $r\in R$.  $(\lambda x.Z_Q)$ thus returns a proof of $Q$, $\neg Q$, or $\bot$, depending on which of these can be constructed given the type of the input argument passed.  We define the probability of a query in terms of the probability that $q_Q$ reduces by $\beta\rho$-reduction to a term $\tau$ of type $Q$ given that it does not reduce to an inconsistency (where $\tau \in \NF_{\lambda P^{\Sigma\BOOL}}$):
\begin{eqnarray}\label{eq:PDTquery4}
P(Q|\Gamma_0,\Gamma_1,P) &=& \frac{P(q_Q\twoheadrightarrow^{\rho}_\beta\tau:Q)}{1-P(q_Q\twoheadrightarrow^{\rho}_\beta\tau:\bot)}.
\end{eqnarray}
The probability in Eq. \ref{eq:PDTquery4} can be evaluated by sampling repeated reductions of Eq. \ref{eq:PDTquery}, and rejecting those samples returning $\tau:\bot$.  Writing $x_W$ for a type $x_W\in\TYPES(D)$, i.e. a possible world, it can be shown (see Appendix, Prop. 8) that:
\begin{eqnarray}\label{eq:PDTquery5}
P(x_W|\Gamma_{0,1},P) &\propto& \sum_{H\in\mathcal{P}(\{1...N_F\})}(\prod_{j\in H} p_j)(\prod_{j\not\in H} (1-p_j))[\forall j\in H, \exists y \st \Gamma^\prime,t:x_W \vdash y:F_j],
\end{eqnarray}
where $[.]$ is the indicator function, which is 1 for a true condition and 0 otherwise, and $\mathcal{P}(.)$ the powerset operator.  Eq. \ref{eq:PDTquery5} implies that the probability for a world under a DTN is proportional to the combined probabilities of all subsets of initial formulae it is consistent with.

We can reformulate the examples from the end of Sec. \ref{subsec:MLN} as DTNs.  For the first, we fix a language via $\Gamma^\prime=<A:*, \bot:*, c_1:A, B^1_1:A\rightarrow *, B^1_2:A\rightarrow *>$, i.e. with one constant and two unary predicates.  We then set the formulae $\Gamma_1 = \Gamma_0, <f_1:(B^1_1(c_1)\rightarrow B^1_2(c_1)), f_2:B^1_1(c_1)>$ with the probabilities $P = \{p_1=(1-e^{-w_1}), p_2=(1-e^{-w_2})\}$.  It can be shown (following the proof of Prop. 1, part (a) below) that the distribution over worlds is identical to that in Sec. \ref{subsec:MLN}, and in particular $P(B^1_2(c_1)|\Gamma_{0,1},P)=\frac{e^{w_1}(1+e^{w_2})}{e^{w_1}(2+e^{w_2})+e^{w_2}}$ as in Eq. \ref{eq:MLNex1} ($w_1,w_2>0$).  For the second example, we let $\Gamma^\prime=<A:*, \bot:*, c_1:A, c_2:A, B^1_1:A\rightarrow *>$, and $\Gamma_1 = \Gamma_0, <f_1:(B^1_1(c_1)), f_2:(B^1_1(c_2)>$, with probabilities $P = \{p_1=p_2=(1-e^{-w_1})\}$.
Again, we can show that the distribution over worlds is equivalent to that in Sec. \ref{subsec:MLN}.  Here, we can write the query $\exists x B^1_1(x)$ using the dependent type $Q = \sum x:A.B^1_1(x)$ (i.e. we desire a pair consisting of $x$, an element of the domain, and a proof that it satisfies predicate $B^1_1$).  As in Eq. \ref{eq:MLNex2}, $P(\sum x:A.B^1_1(x)|\Gamma_{0,1},P)=\frac{e^{w_1}(2+e^{w_1})}{e^{w_1}(2+e^{w_1})+1}$, $(w_1>0)$.

\subsection{Relating MLNs and DTNs}\label{subsec:proof}

We now give a proof of the equivalence of the notions of MLNs and DTNs:

\vspace{0.3cm}
\textbf{Proposition 1.} \textit{The class of distributions over worlds representable by groundings of an MLN is the same as the class of distributions over worlds representable by DTNs.  In particular we have: \\
(a) a canonical translation from ground Markov Networks $M_{L,C}$ to DTNs $\{\Gamma_0,\Gamma_1,P\}$; and \\
(b) a canonical translation from a DTNs to ground Markov Networks, \\
each preserving distributions across worlds.}

\vspace{0.3cm}
\textit{Proof of Proposition 1, part a: }  We provide here a canonical translation from an arbitrary ground MLN $M_{L,C}$ and an equivalent DTN.  We assume that $L$ has been expanded so that any pair $(F,w)$ where $F$ contains free variables has been replaced by $(F_j,w), j=1...J$ with $F_j$ ranging over all possible groundings of the free variables.  We will write $N_F$ for the number of (formula,weight) pairs in $L$, $N_C$ for the number of constants in $C$, and $\mathbf{x}\in\mathbb{B}^P$ for a possible world represented as in Sec. \ref{subsec:MLN} as a binary vector which specifies the truth or falsity of each ground predicate $p=1...P$.

We begin by transforming $L$ to $L^\prime$, where, for pair $(F_i,w_i)\in L$ we set $(F^\prime_i,w^\prime_i)\in L^\prime$ with $F^\prime_i=F_i$, $w^\prime=w$ if $w\leq 0$, and $F^\prime_i=\neg F_i$, $w^\prime=-w$ otherwise.  This transformation will preserve the distribution across worlds, since the ratio $\psi_i(\mathbf{x}_1)/\psi_i(\mathbf{x}_2)$ for a world which satisfies $F_i$, $\mathbf{x}_1$, and one which does not, $\mathbf{x}_2$, is $e^w$ in both networks $M_{L,C}$ and $M_{L^\prime,C}$.

We then construct a DTN by forming $\Gamma_0=\Gamma^\prime,\Gamma^{\prime\prime}$ with the same number of constants, predicates and functions and matching arities as $M_{L,C}$, and setting $\Gamma_1 = \Gamma_0, <f_i:F^{\prime\prime}_i, i=1:N_F>$, for $F^{\prime\prime}_i=\hat{F^\prime_i}\rightarrow\bot$, and $P=\{p_i=(1-\exp(w^\prime_i)), i=1:N_F\}$, where we write $\hat{F}$ for the translation of formulae $F$ into a DTN type expression, using the matching constants, predicates and functions from $\Gamma_0$, and the formulae as types correspondence (discussed in Sec. \ref{subsec:DTN}) to translate the logical symbols into matching type constructors.

We now demonstrate that the DTN constructed above gives the same distribution across possible worlds as $M_{L,C}$.  For a given world $\mathbf{x}$, write $G(\mathbf{x})$ for the proposition $G_1(\mathbf{x})\wedge G_2(\mathbf{x}) \wedge ... G_P(\mathbf{x})$, where for ground predicate $h_p$, $G_p(\mathbf{x}) = h_p$ if $h_p$ is true in world $\mathbf{x}$, and $G_p(\mathbf{x}) = \neg h_p$ otherwise.  Further, let $R_{\mathbf{x}} = \{i\in\{1...N_F\}|G(\mathbf{x})\rightarrow \hat{F^{\prime\prime}_i}\}$ and $S_{\mathbf{x}} = \{i\in\{1...N_F\}|G(\mathbf{x})\rightarrow \neg\hat{F^{\prime\prime}_i}\}$, where $\hat{F^{\prime\prime}_i}$ is the translation of type $F^{\prime\prime}_i$ to a logical formula by the formulae as types correspondence ($R$ is thus the set indices of formulae in $\Gamma_1$ which are consistent with world $\mathbf{x}$, and $S$ the set of indices of formulae that are inconsistent).  Then, from Eq. \ref{eq:PDTquery5}, we have:
\begin{eqnarray}\label{eq:proof1}
P(\hat{G}(\mathbf{x})|\Gamma_{0,1},P) &\propto& \sum_{H\in\mathcal{P}(\{1...N_F\})}(\prod_{j\in H} p_j)(\prod_{j\not\in H} (1-p_j))[\forall j\in H, \exists y \st \nonumber \\
&& \Gamma^\prime,t:\hat{G}(\mathbf{x}) \vdash y:F^{\prime\prime}_j] \nonumber \\
&=& \sum_{r\subset R_{\mathbf{x}}} \prod_{i\in R_{\mathbf{x}}}((p_i)^{[i\in r]}(1-p_i)^{[i\not\in r]})\cdot\prod_{i\in S_{\mathbf{x}}}(1-p_i) \nonumber \\
&=& 1 \cdot \prod_{i\in S_{\mathbf{x}}}(1-p_i) \nonumber \\
&=& \prod_{i\in S_{\mathbf{x}}}\exp(w^\prime_i).
\end{eqnarray}
Similarly, for $M_{L^\prime,C}$ (and thus for $M_{L,C}$ as discussed above) we have:
\begin{eqnarray}\label{eq:proof2}
P(\mathbf{x}|L^\prime,C) &\propto& \prod_i (e^{w^\prime_i})^{[i\in S_{\mathbf{x}}]} \nonumber \\
&=& \prod_{i\in S_{\mathbf{x}}}\exp(w^\prime_i),
\end{eqnarray}
thus giving rise to identical distributions.

\vspace{0.3cm}
\textit{Proof of Proposition 1, part b: } We suppose we have a DTN specified by $\Gamma_0, \Gamma_1, P$, where $\Gamma_1$ contains statements $f_i:F_i$ for $i=1...N_F$.  We use $\mathbf{x}\in\mathbb{B}^P$ as above to represent a possible world, by considering all possible groundings of the predicates in $\Gamma_0$.  Letting $S_{\mathbf{x}} = \{i\in\{1...N_F\}|G(\mathbf{x})\rightarrow \neg\hat{F_i}\}$ (where $\hat{F_i}$ is the translation of type $F_i$ to a logical formula by the formulae as types correspondence), and following the same derivation as Eq. \ref{eq:proof1} lines 1-3, we have:
\begin{eqnarray}\label{eq:proof3}
P(\hat{G}(\mathbf{x})|\Gamma_{0,1},P) &\propto& \prod_{i\in S_{\mathbf{x}}}(1-p_i).
\end{eqnarray}
Consider now a grounded MLN formed using a language containing the same constants, predicates and functions as the DTN above, where we set $L = \{(\neg\hat{F_i},\log(1-p_i)),i=1...N_F\}$.  By Eq. \ref{eq:MLN} we will have
\begin{eqnarray}\label{eq:proof4}
P(\mathbf{x}|L,C) &=& \frac{1}{Z}\prod_i(e^{\log(1-p_i)})^{[i\in S_{\mathbf{x}}]} \nonumber \\
&\propto& \prod_{i\in S_{\mathbf{x}}} (1-p_i),
\end{eqnarray}
hence we have identical distributions.
\begin{flushright}
$\square$
\end{flushright}

Finally, we note that since MLNs provide a universal representation for finite discrete distributions, we have:

\vspace{0.3cm}
\textbf{Corollary 1.} \textit{Both MLNs and DTNs are universal representations for distributions across finitely many discrete variables.}

\vspace{0.3cm}
\textit{Proof.}  \cite{richardson_06} show that MLNs are such a universal representation, and so by Prop. 1 and the fact that each contain only finitely-many discrete variables (/ground predicates), the classes of representable distributions must be all finite discrete distributions for both DTNs and MLNs.  In particular, the translation scheme provided in the proof of Prop. 1 part (a) applied to the canonical MLN representation of a general finite discrete distribution given in \cite{richardson_06} provides a canonical DTN representation of the same distribution.
\begin{flushright}
$\square$
\end{flushright}

\section{Discussion}

We have introduced the probabilistic dependent type system, $\lambda P \rho^{\Sigma\BOOL}$, whose expressions have a two-level semantics denoting distributions across terms and types (or type/type-constructors and kinds), and where $\beta\rho$ reduction corresponds to sampling.  Dependent type networks form a subset of $\lambda P \rho^{\Sigma\BOOL}$, which we have shown to be equivalent in expressive power to Markov Logic Networks, and hence able to represent all distributions over finitely many discrete variables.  The equivalence is provided by a probabilistic analogue of the formulae as types interpretation for typed lambda systems, providing support for the PDTS sampling semantics given (demonstrating broadly that the relationship between $\lambda P \rho^{\Sigma\BOOL}$ and Markov Logic is of the same form as the relationship between $\lambda P^{\Sigma\BOOL}$ and First Order Logic).  In addition the equivalence provides a factor graph semantics (at the type level) for the class of DTNs.

The expressive power of $\lambda P \rho^{\Sigma\BOOL}$ is thus also universal for distributions over finitely many discrete variables, since it contains as a subset DTNs.  A given finite discrete distribution will typically have multiple representations as a PDTS in addition to the representation as a DTN, including ones in which the type and term levels may carry additional semantics relevant to particular application domains.  Further, the sampling based semantics given here suggests that a measure transformer semantics in the sense of \cite{goodman_08} could also be given for $\lambda P \rho^{\Sigma\BOOL}$, which would naturally extend to PDTS over general base types, and allow factor graph message passing algorithms to be used in evaluating expressions.


\appendix

\section{Appendix}\label{app:a}

Proofs of results mentioned in Sec. \ref{sec:lambdaP} and \ref{sec:PDTS} are given below.

\vspace{0.3cm}
\textbf{Proposition 2.} \textit{$\beta$-reduction in $\lambda P^{\Sigma \BOOL}$ preserves types up to $\beta$ equivalence.}

\vspace{0.3cm}
\textit{Proof.}  We first show the proposition folds for all rules in Eq. \ref{eq:betaReduction}.  From the products(1) and products(2) rules in Eq. \ref{eq:typeInf} we have:
\begin{eqnarray}\label{eq:prop2a}
\pi_1 (U_1,U_2)_{\sum x:A.B}:A &\rightarrow_\beta& U_1:A \nonumber \\
\pi_2 (U_1,U_2)_{\sum x:A.B}:B[x:=U_1] &\rightarrow_\beta& U_2:B[x:=\pi_1 (U_1,U_2)_{\sum x:A.B}],
\end{eqnarray}
where, we note that $B[x:=U_1]=_{\beta}B[x:=\pi_1 (U_1,U_2)_{\sum x:A.B}]$.  Further, if we have $\tau_1:B[x=\TRUE]$ and $\tau_2:B[x=\FALSE]$, from the if rule in Eq. \ref{eq:typeInf}:
\begin{eqnarray}\label{eq:prop2b}
(\IF \TRUE \THEN \tau_1 \ELSE \tau_2):B[x=\TRUE] &\rightarrow_\beta& \tau_1:B[x=\TRUE] \nonumber \\
(\IF \FALSE \THEN \tau_1 \ELSE \tau_2):B[x=\FALSE] &\rightarrow_\beta& \tau_2:B[x=\FALSE],
\end{eqnarray}
and from the abstraction and application rules in Eq. \ref{eq:typeInf}, letting $(\lambda x:\tau_1.\tau_2):\prod x:A.B$:
\begin{eqnarray}\label{eq:prop2c}
((\lambda x:\tau_1.\tau_2) \tau_3):B[x:=\tau_3] &\rightarrow_\beta& \tau_2[x:=\tau_3]:B[x:=\tau_3].
\end{eqnarray}

The above cover the base cases of the $\beta$-reduction relation.  For $\tau_1 \rightarrow_\beta \tau_1^\prime$ where $\tau_2 \rightarrow_\beta \tau_2^\prime$, and $\tau_1^\prime$ is the result of replacing free sub-expression $\tau_2$ for $\tau_2^\prime$ in $\tau_1$, we prove the result by induction on the structure of $\tau_1$.  First we outline a general induction step for each form that $\tau_1$ may take.  For $\tau_1 = \prod_ x:A.B:s$, with either the inductive hypothesis $A:*\rightarrow_{\beta}A^\prime:*$ or $B:s\rightarrow_{\beta}B^\prime:s$, we have that $\tau_1^\prime=\prod x:A^\prime.B:s$ or $\tau_1^\prime=\prod x:A.B^\prime:s$ respectively by the type/kind formation rule in Eq. \ref{eq:typeInf} (similarly for $\Sigma$-types).  For $\tau_1=(F a):B[x:=a]$ with inductive hypothesis $F:(\prod x:A.B)\rightarrow_{\beta}F^\prime:(\prod x:A.B)$ or $a:A\rightarrow_{\beta}a^\prime:A$ (where we implicitly use the type conversion rule, Eq. \ref{eq:typeInf}, to equalize types which are $\beta$-equivalent), we have respectively $\tau_1^\prime=F^\prime a:B[x:=a]$ and $\tau_1^\prime=F a^\prime:B[x:=a^\prime]=_{\beta}B[x:=a]$ by the application rule in Eq. \ref{eq:typeInf}.  For $\tau_1=(\lambda x:A.b):(\prod x:A.B)$ and inductive hypothesis $A:s\rightarrow_{\beta}A^\prime:s$ or $b:B\rightarrow_{\beta}b^\prime:B$, we have respectively $\tau_1^\prime=(\lambda x:A^\prime.b):(\prod x:A^\prime.B)=_{\beta}(\prod x:A.B)$ and $\tau_1^\prime=(\lambda x:A.b^\prime):(\prod x:A.B)$ by the abstraction rule in Eq. \ref{eq:typeInf}.  For $\tau_1=\IF a \THEN b_1 \ELSE b_2$ and inductive hypothesis $b_1:B[x:=\TRUE]\rightarrow_{\beta}b_1^\prime:B[x:=\TRUE]$ or $b_2:B[x:=\FALSE]\rightarrow_{\beta}b_2^\prime:B[x:=\FALSE]$, we have respectively $\tau_1^\prime=\IF a \THEN b_1^\prime \ELSE b_2:B[x:=a]$ and $\tau_1^\prime=\IF a \THEN b_1 \ELSE b_2^\prime:B[x:=a]$ by the if rule in Eq. \ref{eq:typeInf}.  Other typing rules in Eq. \ref{eq:typeInf} do not involve sub-expressions, so need not be considered (except the product constant type-tags in the products (1) rule, which can be handled via the type conversion rule).

For the inductive argument, we begin by considering reductions from expressions $\tau_1$ whose maximum sub-expression nesting depth is 1.  Here, $\tau_2$ will correspond to an expression nested directly below $\tau_1$, and thus the inductive hypotheses in all cases considered will be satisfied by one of the base cases in Eqs. \ref{eq:prop2a}, \ref{eq:prop2b} and \ref{eq:prop2c}.  For $\tau_1$ with maximum sub-expression nesting depth $n$ greater than 1, $\tau_2$ will be nested inside a set of sub-expressions of $\tau_1$.  Since these will all have nesting depths less than $n$, the inductive hypothesis can be applied to them if we replace $\tau_2$ by $\tau_2^\prime$, and hence also for $\tau_1$ by one of the cases above.
\begin{flushright}
$\square$
\end{flushright}

\vspace{0.3cm}
\textbf{Proposition 3.} \textit{(Weak-preservation) For expressions $\tau_1$ and $\tau_1^\prime$ in $\lambda P \rho^{\Sigma \BOOL}$ such that $\tau_1 \rightarrow^\rho_\beta \tau_1^\prime$ for non-zero $\rho$, we have that $\TYPES(\tau_1)\supseteq\TYPES(\tau_1^\prime)$, and $\REDUCTIONS(\tau_1)\supseteq\REDUCTIONS(\tau_1^\prime)$, where $\TYPES(\tau_1^\prime)$ and $\REDUCTIONS(\tau_1^\prime)$ are non-empty.}

\vspace{0.3cm}
\textit{Proof.}  We note initially that for any expression $\tau_1$ which is identical to an expression in $\lambda P^{\Sigma \BOOL}$ such that $\tau_1:\tau_2$, by Prop. 2 we have that for $\tau_1 \rightarrow^\rho_\beta \tau_1^\prime$, $\REDUCTIONS(\tau_1)=\REDUCTIONS(\tau_1^\prime)=\{\NORM(\tau_1)\}$ and $\TYPES(\tau_1)=\TYPES(\tau_1^\prime)=\{\NORM(\tau_2)\}$.  For a general expression $\tau_1 \rightarrow^\rho_\beta \tau_1^\prime$, we check as base cases each of the reduction rules in Eq. \ref{eq:betaReduction2}.  The projection rules involve only values, and so are covered by the above.  For $\IF$ reduction, by the if rule in Eq. \ref{eq:typeInf2} we have that for $\REDUCTIONS(\tau_1)=\{b_{1,n}, n = 1...N_{1,1}\}$,$\TYPES(\tau_1)=\{B_{1,n}, n = 1...N_{1,2}\}$,$\REDUCTIONS(\tau_2)=\{b_{2,n}, n = 1...N_{2,1}\}$,$\TYPES(\tau_2)=\{B_{2,n}, n = 1...N_{2,2}\}$:
\begin{eqnarray}\label{eq:prop3a}
\REDUCTIONS(\IF \TRUE \THEN \tau_1 \ELSE \tau_2) &=& \REDUCTIONS(\tau_1) = \{b_{1,n}, n = 1...N_{1,1}\} \nonumber \\
\REDUCTIONS(\IF \FALSE \THEN \tau_1 \ELSE \tau_2) &=& \REDUCTIONS(\tau_2) = \{b_{2,n}, n = 1...N_{2,1}\} \nonumber \\
\TYPES(\IF \TRUE \THEN \tau_1 \ELSE \tau_2) &=& \TYPES(\tau_1) = \{B_{1,n}, n = 1...N_{1,2}\} \nonumber \\
\TYPES(\IF \FALSE \THEN \tau_1 \ELSE \tau_2) &=& \TYPES(\tau_2) =\{B_{2,n}, n = 1...N_{2,2}\}.
\end{eqnarray}
For $\RANDOM$ expressions, by the random and application (1) rules in Eq. \ref{eq:typeInf2} we have for $\REDUCTIONS(F)=\{\lambda x:\BOOL.a_n, n=1...N_1\}$, $\TYPES(F)=\{\prod x:\BOOL.A_n, n=1...N_2\}$:
\begin{eqnarray}\label{eq:prop3b}
\REDUCTIONS(F \TRUE) &=& \{\NORM(a_n[x:=\TRUE]), n=1...N_1\}  \nonumber \\
&\subseteq&\bigcup_{b\in\{\TRUE,\FALSE\}}\{\NORM(a_n[x:=b]), n=1...N_1\}  \nonumber \\&=& \REDUCTIONS(\RANDOM(F)) \nonumber \\
\REDUCTIONS(F \FALSE) &=& \{\NORM(a_n[x:=\FALSE]), n=1...N_1\}  \nonumber \\
&\subseteq&\bigcup_{b\in\{\TRUE,\FALSE\}}\{\NORM(a_n[x:=b]), n=1...N_1\}  \nonumber \\&=& \REDUCTIONS(\RANDOM(F))\nonumber \\
\TYPES(F \TRUE) &=& \{\NORM(A_n[x:=\TRUE]), n=1...N_2\}  \nonumber \\
&\subseteq&\bigcup_{b\in\{\TRUE,\FALSE\}}\{\NORM(A_n[x:=b]), n=1...N_2\}  \nonumber \\&=& \TYPES(\RANDOM(F)) \nonumber \\
\TYPES(F \FALSE) &=& \{\NORM(A_n[x:=\FALSE]), n=1...N_2\}  \nonumber \\
&\subseteq&\bigcup_{b\in\{\TRUE,\FALSE\}}\{\NORM(A_n[x:=b]), n=1...N_2\}  \nonumber \\&=& \TYPES(\RANDOM(F)).
\end{eqnarray}
For $\lambda$ applications of the form $(F a)$ with $F=\lambda x:A.b$, ($b\in\mathcal{T}_x$) and $\Gamma, x:A \vdash \REDUCTIONS(b)=\{b_n, n = 1...N_1\}, \TYPES(b)=\{B_n, n = 1...N_2\}$, where $a:A$, $\REDUCTIONS(a)=\{a\}$, $\TYPES(a)=\{A\}$, $\REDUCTIONS(F)=\{\lambda a:A.b_n, n = 1...N_1\}$, $\TYPES(F)= \{\prod a:A.B_n, n = 1...N_2\}$, we have by the application (1) and abstraction rules in Eq. \ref{eq:typeInf2}:
\begin{eqnarray}\label{eq:prop3d}
\REDUCTIONS(F a) &=& \{\NORM(b_n[x:=a]), n = 1...N_1\}  \nonumber \\
&=& \REDUCTIONS(b[x:=a]) \nonumber \\
\TYPES(F a) &=& \{\NORM(B_n[x:=a]), n = 1...N_2\}  \nonumber \\
&=& \TYPES(b[x:=a]).
\end{eqnarray}
For $\lambda$ applications of the form $(\lambda.Z)a$, where $a:A$, $\REDUCTIONS(a)=\{a\}$, $\TYPES(a)=\{A\}$, and $(A,b)\in\in Z$, $\forall b^\prime \neq b, \neg((A,b^\prime) \in\in Z)$, where $\Gamma, x:A \vdash \REDUCTIONS(b)=\{b_n, n = 1...N_1\}, \TYPES(b)=\{B_n, n = 1...N_2\}$, $b \in \mathcal{T}_x$, we have by the application (1) and (2) rules in Eq. \ref{eq:typeInf2}:
\begin{eqnarray}\label{eq:prop3e}
\REDUCTIONS((\lambda.Z) a) &=& \{\NORM(b_n[x:=a]), n = 1...N_1\}  \nonumber \\
&=& \REDUCTIONS((\lambda x:A.b)a) \nonumber \\
\TYPES((\lambda.Z) a) &=& \{\NORM(B_n[x:=a]), n = 1...N_2\}  \nonumber \\
&=& \TYPES((\lambda x:A.b)a).
\end{eqnarray}

The above form the base cases for the reduction relation $\rightarrow^\rho_\beta$, which show that the proposition holds for each rule in Eq. \ref{eq:betaReduction2} in the case that the sub-expression being reduced is the outermost sub-expression.  For the general case (for a proper sub-expression) we can use a similar inductive argument to Prop. 1 above on the structure on $\tau$, the expression to be reduced, where the induction is on the number of nesting levels $n$ between $\tau_2$, the sub-expression to be reduced, and the outermost level (Eq. \ref{eq:typeInf2} providing all cases for relating sub-expressions at level $n$ to those at level $n-1$).

Further, the $\REDUCTIONS(\tau_1^\prime)$ and $\TYPES(\tau_1^\prime)$ sets after reduction of a legal expression $\tau_1 \rightarrow^\rho_\beta \tau_1^\prime$ must be non-empty, since $\REDUCTIONS(\tau_1^\prime)=\emptyset \Rightarrow \TYPES(\tau_1^\prime)=\{\NOTYPE\}$, and we have $\NOTYPE \not\in \TYPES(\tau_1)$ for legal $\tau_1$.
\begin{flushright}
$\square$
\end{flushright}

\vspace{0.3cm}
\textbf{Proposition 4.} \textit{For any legal expression $\tau_1$ in $\lambda P \rho^{\Sigma \BOOL}$ we have that: (a) (Progress) Either $\tau_1$ is a $\lambda P^{\Sigma \BOOL}$ expression in normal form, or there exists a $\rho$ and $\tau_2$ such that $\tau_1 \rightarrow^\rho_\beta \tau_2$, and that $\sum_{\{(\rho,\tau_2)|\tau_1 \rightarrow^\rho_\beta \tau_2\}}\rho = 1$; and (b) all valid reduction sequences from $\tau_1$ can be extended to a sequence ending in a $\lambda P^{\Sigma \BOOL}$ normal form.}

\vspace{0.3cm}
\textit{Proof.} For part (a) we start by observing that the proposition is already satisfied for terms which are shared between $\lambda P \rho^{\Sigma \BOOL}$ and $\lambda P^{\Sigma \BOOL}$ ($\tau_1 \in \mathcal{T}_{\lambda P^{\Sigma \BOOL}}$) since in $\lambda P^{\Sigma \BOOL}$ we have that leftmost-outermost reduction always finds a normal form by strong normalization (see \cite{thompson_91}, Theorem 5.36).  The proposition automatically follows since there is a unique leftmost-outermost reduction for such terms not in normal form with $\rho = 1$ (the rules in Eq. \ref{eq:betaReduction} become deterministic rules in Eq. \ref{eq:betaReduction2}).

For $\tau_1$ is not in $\mathcal{T}_{\lambda P^{\Sigma \BOOL}}$, we consider first the case that it contains no $\RANDOM_{\rho}(\tau_3)$ sub-expressions.  Then it must contain at least one $(\lambda x.Z)\tau_3$ sub-expression.  For such a sub-expression with the lowest nesting depth, $\tau_3$ will contain no further such sub-expressions, and hence $\tau_3\in\mathcal{T}_{\lambda P^{\Sigma \BOOL}}$, and thus either this expression, or one with a higher nesting level $(\lambda x.Z)\tau_3^\prime$ such that $\tau_3^\prime\in\mathcal{T}_{\lambda P^{\Sigma \BOOL}}$, will be the leftmost-outermost reducible expression of this form, and can be deterministically reduced according to rule 8, Eq. \ref{eq:betaReduction2} (the full conditions of the rule following from the application (2) rule in Eq. \ref{eq:typeInf2} for a legal expression).  Note that sub-expressions of this form are prioritized by condition (i) following Eq. \ref{eq:betaReduction2} above free sub-expressions of other forms. Consider now the case that $\tau_1$ contains at least one $\RANDOM_{\rho^\prime}(\tau_3)$ sub-expression.  Then, we will either have a leftmost-outermost reducible sub-expression of the form $(\lambda x.Z)\tau_3$, or the first $\RANDOM(\tau_3)$ sub-expression encountered by leftmost-outermost order will be available for reduction by rule 5 or 6, Eq. \ref{eq:betaReduction2}, since we placed no further conditions on the reduction of such expressions and they are prioritized over free sub-expressions of other forms (following Eq. \ref{eq:betaReduction2}).  Here, for the two possible reductions by rules 5 and 6, we have $\rho$ values of $\rho^\prime$ and $1-\rho^\prime$, and hence $\sum_{\{(\rho,\tau_2)|\tau_1 \rightarrow^\rho_\beta \tau_2\}}\rho = 1$ as required.

For part (b), we note again that this follows directly when $\tau_1 \in \mathcal{T}_{\lambda P^{\Sigma \BOOL}}$ by strong normalization of $\lambda P^{\Sigma \BOOL}$.  For general terms of $\lambda P \rho^{\Sigma \BOOL}$, part (a) of the proposition ensures that a probabilistic reduction is always possible.  There can be only a finite number of sub-expressions of the form $\RANDOM_{\rho^\prime}(\tau_3)$ and $(\lambda x.Z) \tau_5$ in $\tau_1$ and singly or multiply nested within $\mathcal{Z}$ constructions of $(\lambda x.Z) \tau_5$ expressions within $\tau_1$.  Combining this with the facts that, (1) from condition (i) following Eq. \ref{eq:betaReduction2} (prioritizing the reduction of these expressions); (2) $\RANDOM_{\rho^\prime}(\tau_3)$ expressions are always reducible; (3) $(\lambda x.Z)\tau_3$ expressions are always reducible if no $\RANDOM_{\rho^\prime}(\tau_3^\prime)$ or $(\lambda x.Z)\tau_3^\prime$ expressions are nested beneath them (see part (a) above); and (4) rules 5, 6 and 8 in Eq. \ref{eq:betaReduction2} always decrease the total number of such expressions across all sub-expressions and $\mathcal{Z}$ constructions of $\tau_1$; we can deduce that any valid reduction sequence can be extended to (or already includes as an initial sub-sequence) one which initially reduces all such terms and ends in an expression of $\lambda P^{\Sigma \BOOL}$.  Such a sequence can then be extended to reach a normal form of $\lambda P^{\Sigma \BOOL}$, again by strong normalization.
\begin{flushright}
$\square$
\end{flushright}

\vspace{0.3cm}
\textbf{Proposition 5.} \textit{For legal expressions $\tau_1,\tau_2$ in $\lambda P \rho^{\Sigma \BOOL}$ such that $\tau_1 \twoheadrightarrow^{\rho_0}_\beta \tau_2$, $\rho_0 \leq 1$.}

\vspace{0.3cm}
\textit{Proof.} By Prop. 4, all legal expressions $\tau$ of $\lambda P \rho^{\Sigma \BOOL}$ have at most two valid reductions, and all maximal sequences must terminate in a $\lambda P^{\Sigma \BOOL}$ normal form.  We can thus form a reduction tree for $\tau_1$, with nodes labeled by expressions ($\tau_1$ at the root, the children of a node labeled by the possible reductions of that node, and $\lambda P^{\Sigma \BOOL}$ normal forms at the leaves, repeating expression labels if they are encountered in multiple reductions), and edges labeled by reduction probabilities.  Since by Prop. 4(a) the sum of the edges between a single parent and its children nodes are non-negative and sum to one, writing $\rho_l$ for the product of the weights along the branch from the root leading to leaf $l$, we have that $\sum_l \rho_l = 1$.  Since $\lambda P \rho^{\Sigma \BOOL}$ does not have general recursion, the given expression $\tau_2$ in the proposition can occur at most once along any branch from root to leaf.  Further, we have that for a given node $N$, the product of the weights on the path from the root to $N$ is  $\sum_{l\in L_N} \rho_l$, where $L_N$ contains all leaves whose branches from the root pass through $N$.  Hence, we have $\rho_0 = \sum_{\{l|\exists N (l\in L_N \wedge \psi(N)=\tau_2)\}} \rho_l \leq 0$, where $\psi(N)$ returns the expression with which node $N$ is labeled.
\begin{flushright}
$\square$
\end{flushright}

\vspace{0.3cm}
\textbf{Proposition 6.} \textit{For $\tau_1$ in $\lambda P \rho^{\Sigma \BOOL}$, and any $\tau_3 \in \REDUCTIONS(\tau_1)$, we have that $\tau_1 \twoheadrightarrow^\rho_\beta \tau_3$ with $\rho > 0$, and (a) for each $\tau_3 \in \REDUCTIONS(\tau_1)$, $\exists \tau_4 \in \TYPES(\tau_1) \st \tau_3:\tau_4$; (b) for each $\tau_4 \in \TYPES(\tau_1)$, $\exists \tau_3 \in \REDUCTIONS(\tau_1) \st \tau_3:\tau_4$.}

\vspace{0.3cm}
\textit{Proof.}  We begin by noting that for expressions $\tau_1 \in \mathcal{T}_{\lambda P^{\Sigma \BOOL}}$, the proposition holds, since $\REDUCTIONS(\tau_1)=\{\tau_3\}$ and $\TYPES(\tau_1)=\{\tau_4\}$ are singletons, and we can set $T_{\tau_1}(\tau_3)=\tau_4$ and $R_{\tau_1}(\tau_4)=\tau_3$, where we use $T_{\tau_1}(.)$ to denote a function which, for any $\tau_3^\prime \in \REDUCTIONS(\tau_1)$, picks out a $\tau_4^\prime \in \TYPES(\tau_1)$ for which $\tau_3^\prime:\tau_4^\prime$, and $R_{\tau_1}(.)$ to denote a function which, for any $\tau_4^\prime \in \TYPES(\tau_1)$, picks out a $\tau_3^\prime \in \REDUCTIONS(\tau_1)$ for which $\tau_3^\prime:\tau_4^\prime$.

For the general case, we use an induction on the formation rules for legal expressions in $\lambda P \rho^{\Sigma \BOOL}$, Eq. \ref{eq:typeInf2}, where in each case we assume the proposition holds for the expressions in the antecedent of the rule, and derive from this that it holds for the expression(s) in the consequent.

{\bf(random)}:  For $\tau_1 = \RANDOM_\rho(F)$ we associate with $\tau_3=\NORM(a_n[x:=\TRUE])$ the reduction which first reduces $\tau_1$ to $F \TRUE$, reduces $F \TRUE$ to $(\lambda x:\BOOL.a_n) \TRUE$ (by the inductive hypothesis (IH), noting that leftmost-outermost reduction will reduce expressions in $F$ first), performs the substitution $a_n[x:=\TRUE]$, and finally reduces $a_n[x:=\TRUE]$ to normal form.  Letting $F \twoheadrightarrow^{\rho^\prime}_\beta \lambda x:\BOOL.a_n$, we have $\tau_1 \twoheadrightarrow^{\rho\cdot\rho^\prime}_\beta \tau_3$. Similarly, for $\tau_3=\NORM(a_n[x:=\FALSE])$ we have $\tau_1 \twoheadrightarrow^{(1-\rho)\cdot\rho^\prime}_\beta \tau_3$.  Finally, for $b\in\{\TRUE,\FALSE\}$ we set $T_{\tau_1}(\NORM(a_n[x:=b]))=\NORM(A_m[x:=b])$ where $T_F(\lambda x:\BOOL.a_n)=\prod x:\BOOL.A_m$, and $R_{\tau_1}(\NORM(A_n[x:=b])) = \NORM((R_F(\prod x:\BOOL.A_n))b)$ (for convenience, embedding the meta-language function $R_F(.)$ in the syntax of $\lambda P \rho^{\Sigma \BOOL}$).

{\bf(if)}:  For $\tau_1 = \IF a \THEN a_1 \ELSE a_2$ we associate with $\tau_3=\IF a \THEN b_{1, n_1} \ELSE b_{2,n_2}$ the reduction which reduces $a_1$ to $b_{1,n_1}$, and $a_2$ to $b_{2,n_2}$ (the exact sequence of reduction of sub-expressions will depend on the ordering rules following Eq. \ref{eq:betaReduction2}).  Letting $a_1 \twoheadrightarrow^{\rho_1}_\beta b_{1,n_1}$, and $a_2 \twoheadrightarrow^{\rho_2}_\beta b_{2,n_2}$, we have $\tau_1 \twoheadrightarrow^{\rho_1\cdot\rho_2}_\beta \tau_3$.  Finally, we set $T_{\tau_1}(\IF a \THEN b_{1,n_1} \ELSE b_{2,n_2}) = \IF a \THEN T_{a_1}(b_{1,n_1}) \ELSE T_{a_2}(b_{2,n_2})$ and $R_{\tau_1}(\IF a \THEN B_{1,n_1} \ELSE B_{2,n_2}) = \IF a \THEN R_{a_1}(B_{1,n_1}) \ELSE R_{a_2}(B_{2,n_2})$.

{\bf(abstraction)}:  For $\tau_1 = \lambda x:A.b$ we associate with $\tau_3=\lambda x:A.b_n$ the reduction which prefixes all expressions in a reduction sequence from $b$ to $b_n$ (IH) with $\lambda x:A$, noting that we have that $b\in\mathcal{T}_x$, and so $b$ contains no reducible sub-expressions containing $x$. Letting $b \twoheadrightarrow^{\rho^\prime}_\beta b_n$, we have $\tau_1 \twoheadrightarrow^{\rho^\prime}_\beta \tau_3$.  Finally, we set $T_{\tau_1}(\lambda x:A.b_n) = \prod x:A.T_{b}(b_n)$ and $R_{\tau_1}(\prod x:A.B_n) = \lambda x:A.R_{b}(B_n)$.

{\bf(application (1))}:  For $\tau_1 = F a$ we associate with $\tau_3=\NORM(b_n[x:=a^\prime])$ the reduction which first reduces $F$ to $\lambda x:A.b_n$ with weight $\rho_1$ (IH), then $a$ to $a^\prime$ with weight $\rho_2$ (IH), makes the substitution $b_n[x:=a^\prime]$, and then reduces to normal form.  For this reduction we have $\tau_1 \twoheadrightarrow^{\rho_1\cdot\rho_2}_\beta \tau_3$.  Finally, we set $T_{\tau_1}(\NORM(b_n[x:=a^\prime])) = \NORM(B_m[x:=a^\prime])$ where $T_F(\lambda x:A.b_n)=\prod x:A.B_m$, and $R_{\tau_1}(\NORM(B_n[x:=a^\prime])) = \NORM((R_F(\prod x:A.B_n))a^\prime)$.

{\bf(application (2))}:  For $\tau_1 = (\lambda x.Z) a$ we associate with $\tau_3\in\REDUCTIONS(b[x:=a^\prime])$ the reduction which first reduces $a$ to $a^\prime$ with weight $\rho_1$ (IH), then reduces $(\lambda x.Z) a^\prime$ to $(\lambda x:A.b) a^\prime$ where $a^\prime:A$ and $(A,b)\in\in Z$, and finally reduces $(\lambda x:A.b) a^\prime$ to $b[x:=a^\prime]$ and then to $\tau_3$ with weight $\rho_2$ (IH).  For this reduction we have $\tau_1 \twoheadrightarrow^{\rho_1\cdot\rho_2}_\beta \tau_3$.  Further, we set
$T_{\tau_1}(\tau_3\in\REDUCTIONS(b[x:=a^\prime])) = T_{b[x:=a^\prime]}(\tau_3) \in \TYPES(b[x:=a^\prime])$, and $R_{\tau_1}(\tau_4\in\TYPES(b[x:=a^\prime])) = R_{b[x:=a^\prime]}(\tau_4) \in \REDUCTIONS(b[x:=a^\prime])$.

{\bf($\prod$ / $\sum$ formation)}:  For $\tau_1 = \prod x:A.B$ we associate with $\tau_3=\prod x:A.B_n$ the reduction which prefixes all expressions in a reduction sequence from $B$ to $B_n$ (IH) with $\prod x:A$, noting that we have that $B\in\mathcal{T}_x$, and so $B$ contains no reducible sub-expressions containing $x$. Letting $B \twoheadrightarrow^{\rho^\prime}_\beta B_n$, we have $\tau_1 \twoheadrightarrow^{\rho^\prime}_\beta \tau_3$.  Similarly, by prefixing the same reduction sequence with $\sum x:A$, we have $\tau_1^\prime = \sum x:A.B \twoheadrightarrow^{\rho^\prime}_\beta \tau_3^\prime = \sum x:A.B_n$.  We can further set $T_{\tau_1}(\prod x:A.B_n) = s$, $R_{\tau_1^\prime}(\sum x:A.B_n) = s$, $R_{\tau_1}(s) = \prod x:A.B_1$, and $R_{\tau_1^\prime}(s) = \sum x:A.B_1$.

{\bf(products)}:  For $\tau_1 = \pi_n a$ ($n=1,2$) we associate with $\tau_3=\NORM(\pi_n a^\prime)$ the reduction which first reduces $a$ to $a^\prime$ with weight $\rho^\prime$ (IH), and then reduces $\pi_n a^\prime$ to normal form.  For this reduction we have $\tau_1 \twoheadrightarrow^{\rho^\prime}_\beta \tau_3$.  Further, we set $T_{\pi_1 a}(\NORM(\pi_1 a^\prime))=A$, $R_{\pi_1 a}(A)$ to an arbitrary $\NORM(\pi_1 a^\prime)$, $T_{\pi_2 a}(\NORM(\pi_2 a^\prime))=\NORM(B[x:=\pi_1 a^\prime])$, and $R_{\pi_2 a}(\NORM(B[x:=\pi_1 a^\prime])) = \NORM(\pi_2 a^\prime)$.
\begin{flushright}
$\square$
\end{flushright}

\vspace{0.3cm}
\textbf{Proposition 7.} \textit{(Probabilistic Type Judgment): Letting $\iota_{\tau_1}(\tau_2)=\sum_{\{\tau_3\in\NF_{\lambda P^{\Sigma\BOOL}}|\tau_3:\tau_2\}}\rho_{\tau_3}$, where $\tau_1 \twoheadrightarrow^{\rho_{\tau_3}}_{\beta} \tau_3$, we have (a) $\iota_{\tau_1}(\tau_2)>0\Rightarrow \tau_2\in\TYPES(\tau_1)$ and $\iota_{\tau_1}(\tau_2)=0\Rightarrow \tau_2\not\in\TYPES(\tau_1)$; and (b) $\iota_{\tau_1}(\tau_2)$ can be estimated by the frequency with which $\tau_1$ reduces to $\tau_3\in\NF_{\lambda P^{\Sigma\BOOL}}$ such that $\tau_3:\tau_2$.}

\vspace{0.3cm}
\textit{Proof.} Part (a) follows firstly from that fact that we have shown by Props. 3 and 4b that all maximal reduction sequences of $\tau_1$ end in a single member of $\REDUCTIONS(\tau_1)$ having $\lambda P\rho^{\Sigma \BOOL}$ normal form, and by Prop. 6 that every member of $\REDUCTIONS(\tau_1)$ has a reduction from $\tau_1$ having non-zero probability, implying together that $\REDUCTIONS(\tau_1)$ contains exactly the maximal reductions of all reduction sequences from $\tau_1$; and secondly from the fact that, from Prop. 6 we have that $\tau_2\in\TYPES(\tau_1)\Rightarrow\exists \tau_3\in\REDUCTIONS(\tau_1) \st \tau_3:\tau_2$ and $\tau_3\in\REDUCTIONS(\tau_1)\Rightarrow\exists \tau_2\in\TYPES(\tau_1) \st \tau_3:\tau_2$.  Part (b) follows from the fact that all members of $\REDUCTIONS(\tau_1)$ and $\TYPES(\tau_1)$ are expressions in $\lambda P^{\Sigma \BOOL}$, and that $\beta\rho$ reduction corresponds to sampling from $\REDUCTIONS(\tau_1)$ according to the probabilities $\rho_{\tau_3}$ (which follows from Props. 3a, 5 and 6).  Since the derivability of type judgments (the truth of statements of the form $\tau_3:\tau_2$) is decidable in $\lambda P^{\Sigma \BOOL}$ (see \cite{thompson_91}, Theorem 5.21), we can determine for a given reduction if $\tau_3:\tau_2$, and use the frequency with which this is true to estimate $\iota_{\tau_1}(\tau_2)=\sum_{\{\tau_3\in\NF_{\lambda P^{\Sigma\BOOL}}|\tau_3:\tau_2\}}\rho_{\tau_3}$.
\begin{flushright}
$\square$
\end{flushright}

\vspace{0.3cm}
\textbf{Proposition 8.} \textit{For a Dependent Type Network defined by $\Gamma_0,\Gamma_1$ and $P$ as in Definition 1 (Sec. \ref{subsec:DTN}) and $x_W \in \TYPES(D)$ representing a possible world, with $D$ as in Eq. \ref{eq:PDTquery2}, writing $[.]$ for the indicator function which is 1 for a true condition and 0 otherwise, and $\mathcal{P}(.)$ for the powerset operator, we have:}
\begin{eqnarray}\label{eq:prop6a}
P(x_W|\Gamma_{0,1},P) &\propto& \sum_{H\in\mathcal{P}(\{1...N_F\})}(\prod_{j\in H} p_j)(\prod_{j\not\in H} (1-p_j))[\forall j\in H, \exists y \st \Gamma^\prime,t:x_W \vdash y:F_j].
\end{eqnarray}

\vspace{0.3cm}
\textit{Proof.} By Eq. \ref{eq:PDTquery4} we have that $P(x_W|\Gamma_{0,1},P) \propto P(q_Q \twoheadrightarrow^{\rho}_\beta\tau:Q)$, $Q = x_w$ (where the normalizing constant, $1-P(q_Q\twoheadrightarrow^{\rho}_\beta\tau:\bot) = \sum_W P(x_W|\Gamma_{0,1},P)$, by mutual exclusion of possible worlds, $x_W$).  For $q_Q$ to reduce to $\tau:Q$, two events must occur.  First, $D$ in Eq. \ref{eq:PDTquery} (defined in Eq. \ref{eq:PDTquery2}) must reduce to a term $\tau_1:Q$, since otherwise $\tau_1$ will contain a proof for some ground predicate which is false in $x_W$, allowing a proof of $x_W \rightarrow \bot$, and thus $q_Q$ will reduce to either $\tau:Q\rightarrow\bot$ or $\tau:\bot$.  Writing $N_D$ for the size of the tuple $\tau_1:Q$, by Eq. \ref{eq:PDTquery2} the probability of this occurring is $P_1 = 2^{-N_D}$.  Second, the formulae $F_{i=1...N_F}$ for which the binding of $x_i$ in Eq. \ref{eq:PDTquery} reduces to $f_i:F_i$ must all be consistent with the world represented by $x_W$.  Since $x_W$ fixes all ground predicates, it also fixes all formulae built from these, hence a formula is consistent with $x_W$ if $\hat{x}_W\rightarrow\hat{F_i}$ (where $\hat{x}_W$ is the formulae as types interpretation translation of $x_w$, as discussed in Sec. \ref{subsec:DTN}), or equivalently, $\exists y \st \Gamma^\prime, t:x_W \vdash y:F_i$ ($x_W$ being inhabited implies that $F_i$ is inhabited).  If $x_i$ binds to $f_i:F_i$ for an inconsistent formula, we will be able to construct a proof of $x_W \rightarrow \bot$, and hence again $q_Q$ will reduce to either $\tau:Q\rightarrow\bot$ or $\tau:\bot$.  The probability of this event is $P_2 = \sum_{H\in\mathcal{P}(\{1...N_F\})}P(H)[\forall j\in H, \exists y \st \Gamma^\prime,t:x_W \vdash y:F_j]$, where $P(H)$ is the probability that for the indices $i$ in the subset $H$, $x_i$ is bound to $f_i:F_i$, and for all other formulae $x_i$ is bound to $1:\UNIT$ (hence we sum across all possible subsets of formulae $H$, selecting only those subsets consistent with the query).  Since $P(H) = (\prod_{j\in H} p_j)(\prod_{j\not\in H} (1-p_j))$, we have:
\begin{eqnarray}\label{eq:prop6b}
P(x_W|\Gamma_{0,1},P) &\propto& P_1 \cdot P_2 \nonumber \\
& = & 2^{-N_D} \cdot P_2 \nonumber \\
&\propto& P_2 \nonumber \\
& = & \sum_{H\in\mathcal{P}(\{1...N_F\})}P(H)[\forall j\in H, \exists y \st \Gamma^\prime,t:x_W \vdash y:F_j] \nonumber \\
&=& \sum_{H\in\mathcal{P}(\{1...N_F\})}(\prod_{j\in H} p_j)(\prod_{j\not\in H} (1-p_j))[\forall j\in H, \exists y \st \Gamma^\prime,t:x_W \vdash y:F_j],
\end{eqnarray}
where the step from lines 2 to 3 follows from the fact that $N_D$ is constant for all worlds, and hence $P_1$ may be absorbed into the normalizing constant.
\begin{flushright}
$\square$
\end{flushright}

%
%

\end{document}